\documentclass[lettersize,journal]{IEEEtran}
\usepackage{amsmath,amsfonts,amsthm}
\usepackage{bbm} % 黑板粗体。注意：若只要 \mathbb，amssymb 也能满足大部分需求

\usepackage{algorithm}
\usepackage{algorithmic}
\usepackage{graphicx}
\usepackage{wrapfig}
\usepackage{float}
\usepackage{stfloats}
\usepackage{tabularx}
\usepackage{array}
\usepackage{booktabs}
\usepackage{multirow}
\usepackage{longtable}
\usepackage{tabu}
\usepackage{threeparttable}
\usepackage{tablefootnote}
\usepackage{makecell}
\usepackage{adjustbox}
\usepackage{arydshln}
\usepackage{colortbl}
\usepackage[caption=false,font=normalsize,labelfont=sf,textfont=sf]{subfig}
\usepackage{enumitem}
\usepackage{textcomp}
\usepackage{romannum}
\usepackage{xspace}
\usepackage{pifont}
\usepackage{soul}
\usepackage[normalem]{ulem}
\usepackage{setspace}
\usepackage{verbatim}

\usepackage{xcolor}
\usepackage{tcolorbox}
\usepackage{url}
\usepackage[numbers,sort&compress]{natbib}
\usepackage{hyperref} % hyperref 放在 natbib 后面
\usepackage{lipsum}
\usepackage[utf8]{inputenc}
\def\name{{StriderSPD}\xspace}
\definecolor{background}{rgb}{0.94, 0.97, 1.0}
\definecolor{edge}{rgb}{0.32, 0.48, 0.72}

\begin{document}

% \title{Structure-Guided Joint Representation Learning of Binary Code for Security Patch Detection}
\title{\name: Structure-Guided Joint Representation Learning for Binary Security Patch Detection}

\author{Qingyuan Li, Chenchen Yu, Chuanyi Li, Xin-Cheng Wen, Cheryl Lee, Cuiyun Gao, and Bin Luo
% ~\IEEEmembership{National Key Laboratory for Novel Software and Technology,~Nanjing University,}

\thanks{This research/project is supported by National Natural Science Foundation of China (62572235) and CCF-Huawei Populus Grove Fund. Any opinions, findings, and conclusions or recommendations expressed in this material are those of the author(s). Chuanyi Li is the corresponding author.}% <-this % stops a space
 \thanks{Qingyuan Li, Chenchen Yu, Chuanyi Li, and Bin Luo are with the National Key Laboratory for Novel Software Technology, Nanjing University (Emails: \{522022320072,211250174\}@smail.nju.edu.cn, \{lcy,luobin\}@nju.edu.cn).}
\thanks{Xin-Cheng Wen and Cuiyun Gao are with the Harbin Institute Of Technology Shenzhen (Emails: xiamenwxc@foxmail.com, gaocuiyun@hit.edu.cn).}
\thanks{Cheryl Lee is with the Chinese University of Hong Kong (Email: cheryllee@link.cuhk.edu.hk).}}

\markboth{IEEE Transactions on Software Engineering,~Vol.~14, No.~8, August~2026}%
{Li \MakeLowercase{\textit{et al.}}: XXX}


\maketitle

\begin{abstract}
% Suggestions from Cheryllee: 压缩背景 + 有侧重地提出challenge以平衡叙述占比 + 宏观地提出我们的方法并一一回扣到challenge上 + 层次化、递进的challenge
%%%% 大背景：漏洞危害
Vulnerabilities severely threaten software systems, making the timely application of security patches crucial for mitigating attacks.
%%%% 小背景：SPD的重要性
However, software vendors often silently patch vulnerabilities with limited disclosure, where Security Patch Detection (SPD) comes to protect software assets.
%%%% 明确的场景限定
Recently, most SPD studies have targeted Open-Source Software (OSS), yet a large portion of real-world software is closed-source, where patches are distributed as binaries without accessible source code. % lqy: 这里更适合用现在完成时，暗含对闭源场景的持续忽略
%%%% 一大一小challenge
The limited binary SPD approaches often lift binaries to abstraction levels, i.e., assembly code or pseudo-code.
However, assembly code is register-based instructions conveying limited semantics, while pseudo-code lacks parser-compatible grammar to extract structure, both hindering accurate vulnerability-fix representation learning. In addition, previous studies often obtain training and testing data from the same project for evaluation, which fails to reflect closed-source conditions.
%%%% 宏观的对方法的介绍
To alleviate the above challenges, we propose \textbf{\textit{\name}}, a \underline{Str}ucture-gu\underline{ide}d joint \underline{r}epresentation \underline{SPD} framework of binary code that integrates a graph branch into a large language model (LLM), leveraging structural information to guide the LLM in identifying security patches. 
%%%% 我们提出的方法实现过程中的challenge（与上文challenge呈递进关系）
Our novel design of the adapters in the graph branch effectively aligns the representations between assembly code and pseudo-code at the LLM's token level. 
We further present a two-stage training strategy to address the optimization imbalance caused by the large parameter disparity between \name's two branches, which enables proper branch fitting.
%%%% 解决场景失真
To enable more realistic evaluation, we construct a binary SPD benchmark that is disjoint from prior datasets in both projects and domains and extensively evaluate \name on this benchmark.
% 实验1结果
Experimental results demonstrate the effectiveness of \name, surpassing the best-performing baseline by \textbf{12.66\%} in accuracy.
% 实验3结果
Furthermore, \name exhibits generalizability to different LLMs, enhancing the performance of Llama, Qwen, and DeepSeek on binary SPD tasks.
\end{abstract}


\section{Introduction} \label{section:intro}

%%%% General background
\IEEEPARstart{S}{oftware} systems have undergone an unprecedented expansion, giving an increasing number and diversity of vulnerabilities that pose severe risks to software security and society~\cite{telang2007empirical}.
According to the 2025 OSSRA Report, 86\% of codebases contained at least one open-source vulnerability, with an average of 154 unique vulnerabilities per codebase~\cite{OssraReport}. Furthermore, the Economic Impact of Cybercrime Report in 2018 estimates that the global cost of cybercrime is close to 600 billion dollars~\cite{EICReport}.
%% Topic background
However, software vendors often release security patches without sufficient disclosure (e.g., Common Vulnerabilities and Exposures)~\cite{li2017large}.
%% Topic importance
As a result, timely detection and update of security patches is critical for defeating potential attacks. 

%%%% Binary SPD最直接的challenge
Despite numerous Security Patch Detection (SPD) methods that have been proposed for Open-Source Software (OSS), real-world ecosystems largely consist of closed-source applications and proprietary systems, where patches are released as binaries. As compilation strips away both semantic and structural information~\cite{linn2003obfuscation}, rendering raw-byte analysis impractical.
%%%% Existing methods
Given that assembly code and pseudo-code are the most typical abstraction levels obtained when reversing native binaries~\cite{cifuentes1995decompilation}, the limited binary SPD methods utilize either of them to identify security patches~\cite{li2025empiricalstudycodelarge,he2024bingo,peng20191dvul,xu2017spain}.
%% assembly-code-based methods
Most of them leverage assembly code, as disassembly techniques are relatively straightforward and well established.
For example, SPAIN~\cite{xu2017spain} detects changed basic blocks in the Control Flow Graphs (CFGs) of assembly code and performs semantic analysis on these changed traces to identify security patches. 1dVul~\cite{peng20191dvul} identifies target branches in assembly code using heuristic rules and applies a symbolic execution to generate inputs that reach the target branch and verify the security relevance of patches. BinGo~\cite{he2024bingo} employs a Graph Neural Network (GNN) to represent code property graphs derived from CFGs for security-patch identification.
%% pseudo-code-based methods
Conversely, \citet{li2025empiricalstudycodelarge} conducted a comprehensive comparison between assembly code and pseudo-code. Their empirical study indicates that pseudo-code is better suited for binary SPD tasks when utilizing Large Language Models (LLMs) despite the relative complexity of decompilation techniques.

%%%% Limitations
However, all existing binary SPD methods, whether utilizing assembly code or pseudo-code, suffer from inherent limitations. Methods that leverage assembly code are generally effective at capturing structural information but remain limited in representing semantic information, as assembly code inherently consists of low-level, register-based machine instructions that lack high-level semantics~\cite{kruegel2004static,cifuentes1995decompilation,cooper2002building}. 
In contrast, methods utilizing pseudo-code capture semantic information well but are less effective at modeling structural information. This is because pseudo-code lacks a formal, parser-compatible grammar for extracting the program structure (e.g., CFG or AST)~\cite{avgerinos2011tie,banerjee2021variable}, and the serialization of pseudo-code inevitably discards part of the code hierarchy, exacerbating this limitation.
%% Limitations -> Challenge1
These limitations pose a serious challenge to simultaneously capturing structural and semantic information for effectively representing binary patches, as prior studies have shown that modeling structural information can enhance the understanding of code semantics~\cite{guographcodebert,guo2022unixcoder}.
%% Challenge2
Furthermore, existing binary SPD methods are typically evaluated under unrealistic conditions~\cite{he2024bingo,peng20191dvul,xu2017spain}, undermining the validity of their assessment.

%%%% Challenge1 -> Designing1
Motivated by the challenge of effectively capturing both structural and semantic information, we propose \textbf{\name}, a structure-guided binary SPD framework that jointly represents assembly code and pseudo-code.
\name comprises two key components:
(1) A structure-guided joint representation neural network that integrates a graph branch into an LLM branch, which provides structural information to the LLM at a finer-grained level. This neural network enables the graph branch to align the graph representation with the LLM's latent embedding via adapter-mediated alignment. Specifically, the graph branch employs a GNN to represent pre- and post-patch assembly-code CFGs, subsequently utilizes three adapters to map the graph representation into the LLM's latent space corresponding to queries (Q), keys (K), and values (V), enabling a tighter integration between structural and semantic information, and finally performs cross-attention between the adapted graph representation and the output of the LLM's final feed-forward layer, aligning structural information with semantic information at the sequence level. The LLM branch employs Qwen3-8B~\cite{yang2025qwen3} as the backbone to embed the instruction that contains the pre- and post-patch pseudo-code functions. 
(2) A two-stage training strategy, which mitigates the optimization imbalance caused by the significant disparity in parameter scale between the two branches. Specifically, in the first stage, the LLM is fine-tuned via supervised instruction tuning, enabling it to follow the instruction, capture semantic information from pseudo-code, and effectively separate positive from negative samples in its latent space. In the second stage, the LLM parameters are frozen, and only the graph branch is trained. Anchored by the LLM, the graph branch is driven to align structural and node-level information from assembly code with the LLM’s latent embedding.
%%%% Challenge2 -> Designing2
To further mitigate the challenge posed by unrealistic evaluation conditions, we construct a large-scale benchmark derived from projects and domains entirely distinct from those of prior binary SPD datasets~\cite{li2025empiricalstudycodelarge,he2024bingo}. Our benchmark comprises \textbf{1,720} binary patches across five optimization levels (O0, O1, O2, O3, and Os).

%%%% Evaluation
We evaluate \name against seven binary SPD baselines and two source-code baselines adapted to binary patches. Additionally, we include two state-of-the-art (SOTA) proprietary LLMs for comparison.
%%%% Evaluation results
%% 主试验结果
Experimental results demonstrate that \name outperforms the best baseline Yi-Coder-9B-Chat~\cite{yi-coder}, achieving improvements of \textbf{12.66\%}, \textbf{8.20\%}, and \textbf{38.57\%} in terms of accuracy, F1 score, and false positive rate, respectively.
%% 不同类型补丁识别结果
It also achieves SOTA identification accuracy of \textbf{0.935} on security patches for buffer overflow, the most prevalent vulnerability type, surpassing Yi-Coder-9B-Chat by \textbf{5.65\%}.
%% 泛化性实验结果
Furthermore, \name exhibits strong generalizability across different foundation language models, enhancing the binary SPD capabilities of Llama~\cite{touvron2023llama}, Qwen~\cite{qwen2}, and DeepSeek~\cite{deepseekllm}. On the most compatible foundation LLM, Qwen3-8B, it achieves substantial improvements of \textbf{32.81\%} in accuracy, \textbf{25.53\%} in F1 score, and \textbf{38.32\%} in false positive rate.
%% 强调效率与可用性
% \lqy{
Beyond its superior performance, \name is highly practical for real-world deployment. 
It requires only \textbf{0.67s} on average for inference, and its total cost is just \textbf{\$21.62}, substantially lower than the \$74 associated with GPT-4o.
Its variant built upon Qwen2.5-Coder-0.5B-Instruct boosts the base model’s performance, achieving improvements of \textbf{11\%} in accuracy and \textbf{9.8\%} in F1 score, demonstrating its potential for resource-constrained scenarios.
% }

%%%% Contributions
In summary, the main contributions of this paper are as follows:
\begin{itemize}
    % Methodology
    \item To the best of our knowledge, we propose \name, the first structure-guided joint representation SPD framework of binary code, enabling simultaneous capture of structural and semantic information for more effective security patch detection.
    
    % Benchmark
    \item We construct a cross-project and cross-domain benchmark to more faithfully reflect real-world closed-source scenarios, where historical security patch data from the same project is inaccessible.
    
    % Evaluation
    \item We conduct an extensive evaluation of \name against 11 baselines, and further investigate the generalizability of \name, demonstrating that \name outperforms the best-performing baseline and performs effectively across multiple widely used foundation language models.
\end{itemize}
\section{Preliminary} \label{section:pre}

\subsection{Problem Statement}
In closed-source software, SPD aims to determine whether the differences between a pre-patch binary and its post-patch counterpart indicate a vulnerability fix. 
Formally, given a pre-patch binary $bin_{pre}$ and its post-patch counterpart $bin_{pos}$, the task aims to design a classifier that determines whether the changes from $bin_{pre}$ to $bin_{pos}$ correspond to a security patch. 
% \lqy{
However, vulnerability-fixing changes are often entangled with non-security modifications within the same patch between two binary versions~\cite{arakawa2025towards,luo2024strengthening,sun2025dispatch}. To avoid losing change meaning (too fine-grained) or label noise (too coarse-grained), we follow prior work~\cite{lang2021pmatch,xu2017spain,peng20191dvul} to perform binary SPD at the function level by classifying each changed function pair.
Accordingly, we formalize binary SPD as shown in Formula~\ref{equation:taskdef}.
% }
\begin{equation}\label{equation:taskdef}
\begin{aligned}
    \mathit{Classifier}(\langle f_i^{pre}, f_i^{post} \rangle) &= \{ \textit{security} \mid \textit{non-security} \} \\
    % \langle f_i^{pre}, f_i^{post} \rangle &= f_i\\
    \langle f_i^{pre}, f_i^{post} \rangle  &\in \mathit{diff}(\mathit{bin}^{pre}, \mathit{bin}^{post})
\end{aligned}
\end{equation}
In this formula, \textit{Classifier}() is the classifier to be designed, $\langle f_i^{pre}, f_i^{post} \rangle$ represents a changed function pair from pre- and post-patch binaries, and \textit{diff}() denotes a binary diffing tool that extracts matched pairs of changed functions, such as BinDiff\footnote{\url{https://github.com/google/bindiff}} and DeepBinDiff\footnote{\url{https://github.com/yueduan/DeepBinDiff}}. 
Note that this paper focuses on detecting security-related changes, while binary diffing tools are beyond the scope of our study. 

\begin{figure*}[htbp]
\centering
\includegraphics[width=0.95\textwidth]{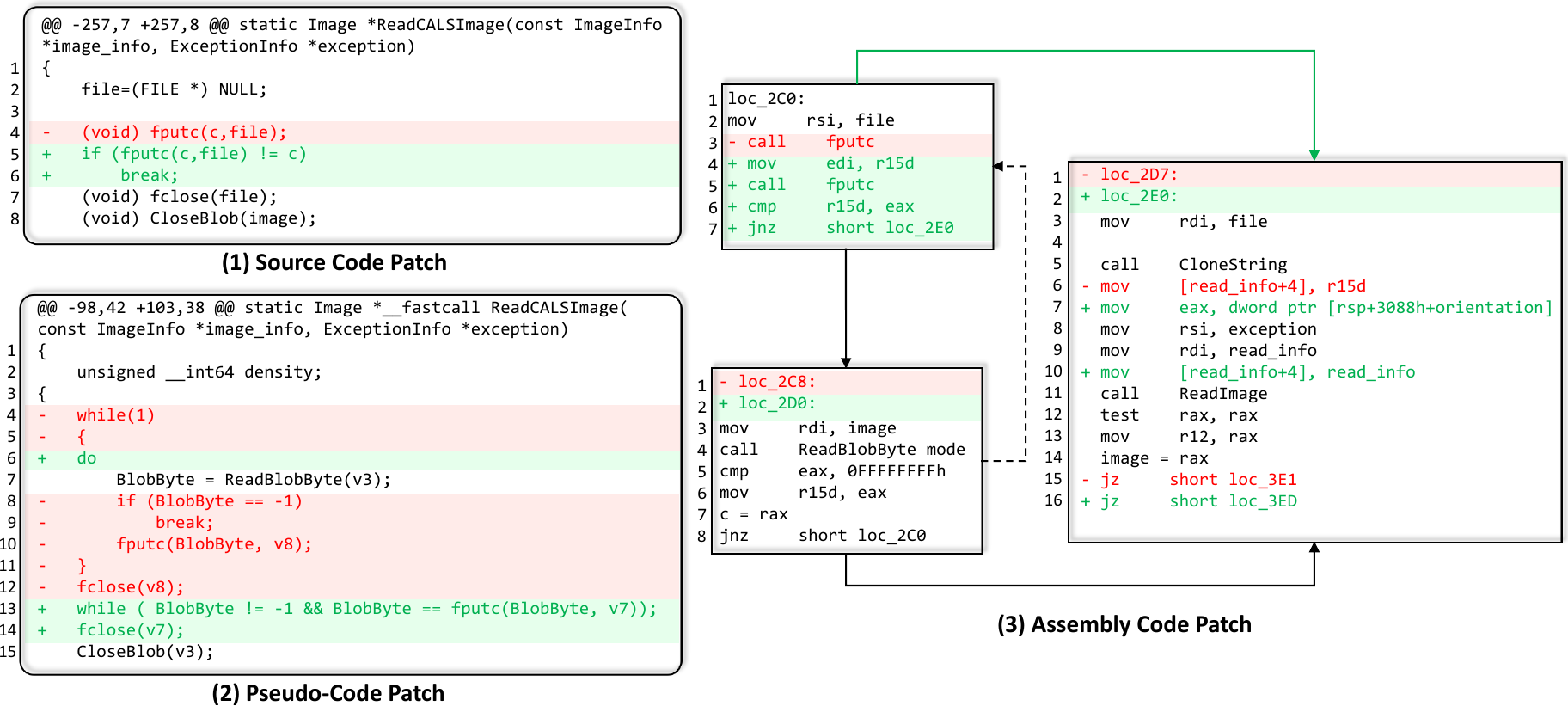}
\vspace{-0.5em}
\caption{Motivating example of the patch fixing CVE-2018-16643.}
\label{fig:motivation}
\vspace{-1em}
\end{figure*}

\subsection{Motivation}\label{section:motivation}

%%%% 场景限定
In the real world, many software systems are closed-source or proprietary~\cite{economides2006two,lee2005open}. To determine whether a patch to such a system fixes a security vulnerability, the only way is to analyze the pre- and post-patch binary files. 
%%%% 直观的方法是直接在raw byte上识别SPD
Since compilation removes both semantic and structural information from machine code~\cite{linn2003obfuscation}, existing binary SPD methods typically leverage higher abstraction levels, such as assembly code or pseudo-code, to analyze binaries.
%%%% 现有工作的limitation
However, these methods suffer from inherent limitations, as they rely solely on either assembly code or pseudo-code, thereby capturing only structural or semantic information effectively.
%%%% 介绍assembly，pseudo-code
Specifically, assembly code is organized into basic blocks linked via control-transfer instructions (e.g., unconditional jump, conditional branches, and call/return), naturally forming a graph that encodes structural information about the program's control flow~\cite{cooper2002building,brumley2013native}, but consisting of register-based instructions that lack high-level semantics~\cite{kruegel2004static,cifuentes1995decompilation,cooper2002building}. 
In contrast, pseudo-code recovers high-level constructs (e.g., local variables and expressions), yielding a human-readable C-like code that closely resembles source code and conveys more semantic information~\cite{brumley2013native,burk2022decomperson}, but lacks a parser-compatible grammar, hindering the extraction of structural information~\cite{avgerinos2011tie,banerjee2021variable}.
% lqy: edited at 09.12-09:30
Taken together, since structural and semantic information cannot both be captured effectively at a single level of abstraction, relying on only one level is insufficient to accurately identify security patches.

%%%% 先给motivating example
Figure~\ref{fig:motivation} is a real-world example patch that addresses an unchecked return value vulnerability identified as CVE-2018-16643~\cite{cve201816643} in ImageMagick’s coders/pict.c.
The vulnerability arises because the return value of the \textit{fputc} function is not verified in line 4 of Figure~\ref{fig:motivation}(1), thereby allowing remote attackers to trigger a denial-of-service attack via crafted image files. In lines 5 and 6 of Figure~\ref{fig:motivation}(1), the source-code patch introduces a check on the return value of the \textit{fputc} function and immediately breaks the loop when \textit{fputc(c, file) != c}, thereby providing both semantic information (the if-statement predicate) and structural information (an if-triggered control-flow exit) necessary to comprehend this security fix.

In the pseudo-code patch, the if condition in the source code is decompiled into a control-flow transformation, replacing the \textit{while} loop in line 4 of Figure~\ref{fig:motivation}(2) together with the \textit{if-break} in lines 8–12, by a \textit{do-while} loop spanning lines 6 and 13–14. Since pseudo-code is only available to an LLM in serialized form, both constructs appear as loop-termination conditions in the sequence. Therefore, they may be interpreted as the same paradigm, making it difficult to determine whether the change introduces an early exit to prevent the vulnerability or merely rewrites the loop termination condition. % lqy: 这样表述是不是就把1.pseudo-code难以抽取结构特征 + 2.LLM普遍难以利用结构化信息融合起来了？
In contrast, the assembly patch preserves structural information but is harder to interpret semantically.
Specifically, in Figure~\ref{fig:motivation}(3), lines 4 and 5 of the first basic block \textit{loc\_2C0} introduce the instructions \textit{cmp r15d, eax} and \textit{jnz short loc\_2E0}, which add an edge from \textit{loc\_2C0} to the third basic block \textit{loc\_2E0} and simultaneously modify the instructions within the basic blocks, thereby altering both the CFG topology and the node embeddings.
These patterns of structural and node-level changes, such as early exits and guard nodes, are often key indicators of security patches, and GNNs are particularly effective at capturing these changes. However, compared with the conditional statements in source code, register instructions such as \textit{cmp r15d, eax} are more difficult to comprehend.

%%%% Key Motivation: 同时利用assembly和pseudo
The example shows that assembly code preserves structural information well, while pseudo-code captures semantic information well. Motivated by this, we aim to jointly represent assembly-code structures and pseudo-code semantics, enabling models to better comprehend binaries and thereby accurately detect security patches.
% %%%% Key Motivation: 同时利用assembly和pseudo
% Our motivation is to jointly represent assembly-level structures and pseudo-code semantics, enabling models to better comprehend binaries and thereby accurately detect security patches.
%%%% Motivation 1: 应该利用LLM的能力在code-related task上，直观的方法是把所有表示都丢进去。
A straightforward idea is to feed LLMs with assembly code and pseudo-code simultaneously, given that LLMs have revolutionized various software engineering tasks and demonstrated promising results for binary-related tasks~\cite{tan2024llm4decompile,cummins2025llm}.
% A straightforward idea is to feed LLMs with assembly code and pseudo-code simultaneously, enabling them to capture structural and semantic information from joint representations, given that LLMs have revolutionized various software engineering tasks and demonstrated promising results for binary-related tasks~\cite{tan2024llm4decompile,cummins2025llm}.
%%%% Motivation 2: 进一步地说明LLM不能直接序列化assembly以得到结构信息
However, naively serializing low-level code (e.g., assembly or compiler IR) into a flat token sequence as input to LLMs fails to capture such structural information~\cite{niu2024fair}. 
First, the pretraining corpora of most LLMs are dominated by source code rather than low-level code~\cite{yi-coder,deepseekllm,deepseekcoderv2,qwen2,qwen2.5coder,yang2025qwen3}, so the high-level semantics learned by LLMs are inherently misaligned with assembly’s low-level semantics, causing LLMs to produce confused representations when applied to assembly-related tasks~\cite{chen2023investigating}.
Second, LLMs are typically standard Transformer-based architectures~\cite{NIPS2017_3f5ee243}, optimized for sequential modeling. They cannot natively encode graphs to capture structural information as effectively as structure-aware architectures~\cite{wu2021representing}.

%%%% Behavior guided by motivations: 既利用assembly-code的结构信息，又利用LLM强大的语义理解，但却不影响LLM
To overcome the inability of LLMs to capture structural information, we propose a binary SPD framework named \name, which (1) enables LLMs to leverage the structural information in assembly-code CFGs without being confused by the low-level semantics, and (2) leverages LLMs’ remarkable capabilities of code semantic understanding to capture the semantics of pseudo-code, thereby enabling accurate security patch identification.

\section{Benchmark Construction} \label{section:bench}

% 真实场景的挑战造成了闭源补丁安全相关的不可知。
It is challenging to determine whether patches in real-world closed-source software are security-related. Existing datasets typically simulate closed-source environments by compiling patches from OSS~\cite{he2024bingo,li2025empiricalstudycodelarge}, using raw data drawn from the same source-code repositories. These approaches often partition the data into training and test sets at the patch level, resulting in patches from the same project appearing in both sets. Such a setup fails to accurately evaluate SPD methods under realistic closed-source conditions.
% 跨项目甚至跨领域最能模拟真实的闭源场景。
Given the inability to reliably determine whether historical changes in closed-source software correspond to vulnerability fixes, datasets used for training and testing SPD models should be drawn from entirely distinct projects. Ensuring complete disjointness between training and test sets across projects is essential for a rigorous assessment of real-world closed-source SPD scenarios.
In this paper, we construct a large-scale benchmark to evaluate the effectiveness of binary SPD methods in real-world closed-source SPD scenarios. The benchmark is constructed as follows.

\textbf{\textit{Projects Selection}}:
We first identify the distribution of projects in the publicly released binary SPD dataset~\cite{li2025empiricalstudycodelarge} and determine the domains to which these projects belong. 
Subsequently, following the prior studies~\cite{he2024bingo,li2025empiricalstudycodelarge}, we extract projects from two popular source-code patch datasets, ReposVul~\cite{wang2024reposvul} and PatchDB~\cite{wang2021patchdb}, that are not included in the existing dataset. %lqy: 引用上一篇empirical
We further filter out projects that do not belong to the same domains as those in the existing dataset. 
Finally, from these projects, we select five high-quality projects: ImageMagick~\cite{ImageMagick}, TcpDump~\cite{TcpDump}, Qemu~\cite{Qemu}, Radare2~\cite{Radare2}, and Slurm~\cite{Slurm}, each with over 3,000 GitHub stars, as the project sources for constructing our benchmark.

\textbf{\textit{Raw Data Quality Inspection}}: 
% \lqy{
We manually inspect 1,068 source code files to be compiled. We assign a sample a \textit{Positive} label only when its CVE identifier can be unambiguously matched to the corresponding repository and commit in the metadata; otherwise, samples without a CVE identifier are labeled as \textit{Negative}.
% }

\textbf{\textit{Compilation and Decompilation}}: 
% \lqy{
Following previous studies~\cite{he2024bingo,li2025empiricalstudycodelarge}, we compile these source code files with the GCC compiler at five optimization levels (O0, O1, O2, O3, and Os) to generate binary files. 
We then decompile the resulting binaries using IDA Pro\footnote{\url{https://hex-rays.com/ida-pro}}, a widely used decompiler with remarkable decompilation robustness~\cite{cao2024evaluating,liu2020far}. This process produces lifted files in both assembly and pseudo-code representations, achieving a decompilation success rate of 74\%.
Overall, the benchmark comprises 1,720 functions, with 1,010 positive and 710 negative samples, providing a more realistic foundation for evaluating security patch detection methods.
% }

Table~\ref{tab:bench_info} presents the detailed project sources, project domains, and data distributions across the projects for both the dataset from the prior study~\cite{li2025empiricalstudycodelarge} and our benchmark.
\begin{table}[t!]
\centering
\caption{Project information of the dataset and our benchmark}
\label{tab:bench_info}
\resizebox{0.48\textwidth}{!}{
\begin{tabular}{ccccc}
\toprule
\textbf{Source} & \textbf{Project} & \textbf{Domain} & \textbf{Star} & \textbf{Distribution} \\
\cmidrule{1-5}\morecmidrules\cmidrule{1-5}

% \multirow{5}{*}{\makecell[c]{\textit{\protect\citet{li2025empiricalstudycodelarge}'s} \\ \textit{Dataset}}}
\multirow{5}{*}{\makecell[c]{\textit{Li et al.'s}\\ \textit{Dataset}}}
& Linux~\cite{Linux} & Operating System & 201k & 90.8\% \\
& FFmpeg~\cite{FFmpeg} & Multimedia & 52.6k & 5.7\% \\
& Git~\cite{Git} & Version Control & 56.2k &  1.7\% \\
& Php~\cite{Php} & Web & 39.4k &  1.2\% \\
& Libav~\cite{Libav} & Multimedia & 1.1k & 0.7\% \\
\midrule

\multirow{5}{*}{\makecell[c]{\textit{Our}\\ \textit{Benchmark}}}
& ImageMagick~\cite{ImageMagick} & Image & 14.3k &  31.7\% \\
& TcpDump~\cite{TcpDump} & Network & 3k & 27.0\% \\
& Qemu~\cite{Qemu} & Virtualization & 11.9k &  30.5\% \\
& Radare2~\cite{Radare2} & Security & 22.1k &  7.8\% \\
& Slurm~\cite{Slurm} & Cluster & 3.2k &  2.9\% \\
\bottomrule
\end{tabular}
}
\end{table}

\section{Methodology} \label{section:method}

\begin{figure*}[t!]
\centering
\includegraphics[width=0.85\textwidth]{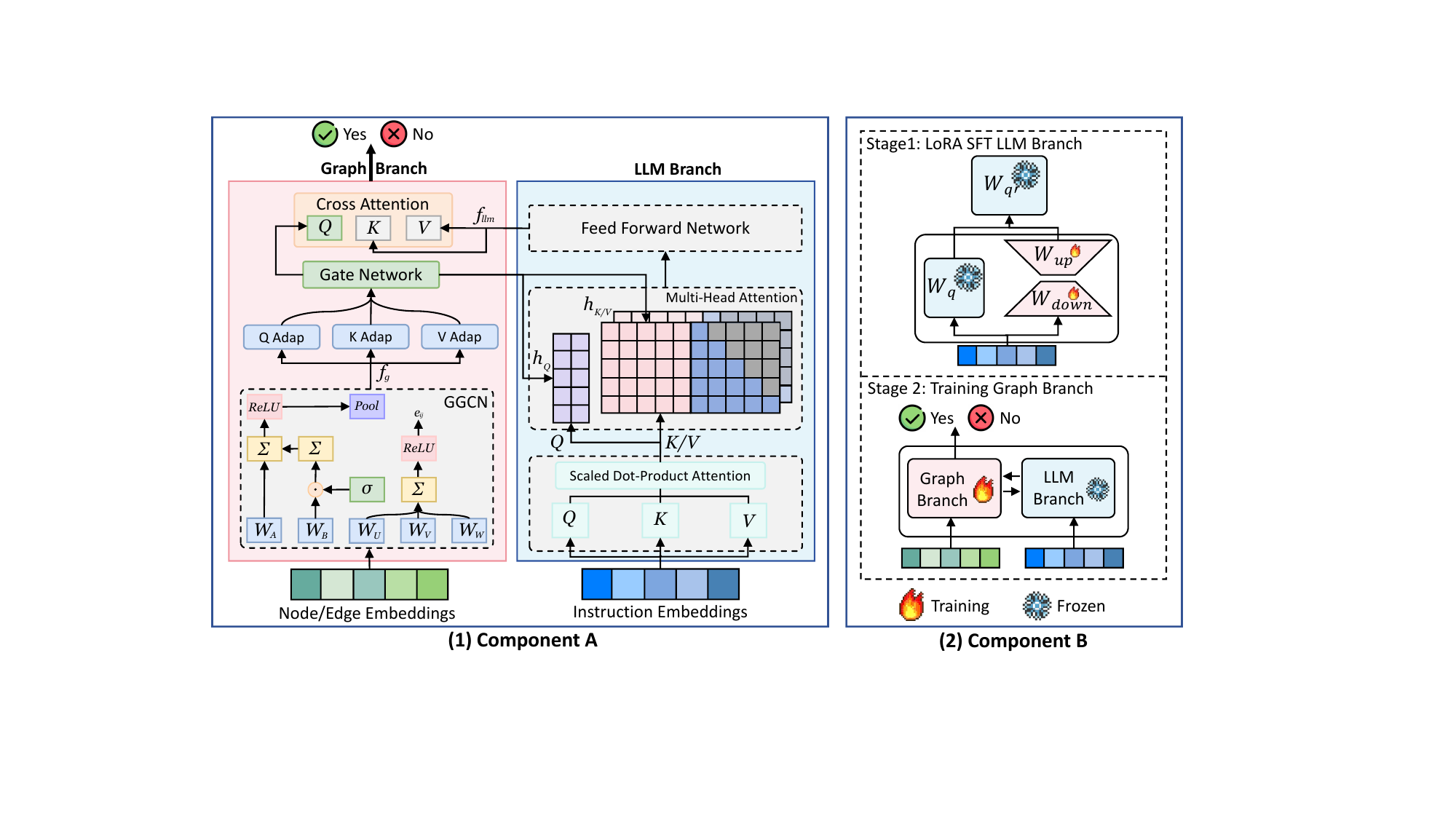}
\vspace{-0.5em}
\caption{Overview of \name.}
\label{fig:overview}
\vspace{-1em}
\end{figure*}

Figure~\ref{fig:overview} provides an overview of \name, which comprises two main components:
% . It mainly comprises two components: 
(1) a structure-guided neural network, which integrates an LLM branch and a graph branch to construct a joint representation of assembly code and pseudo-code, and (2) a two-stage training strategy, which mitigates the optimization imbalance caused by the significant disparity in parameter scale between the two branches and accelerates the training process.

\subsection{Structure-Guided Joint Representation} \label{section:arch} 

\name employs a graph branch to encode the structural information from assembly-code patches, integrates the graph representations into the LLM branch, and utilizes the LLM to capture semantics of pseudo-code patches.
% \lqy{
Given prior work that CFGs are one of the most effective and widely adopted representations for capturing program topological structure~\cite{xu2017neural,wang2024improving}, \name uses the assembly-function CFG as input to the graph branch. A CFG's nodes correspond to basic blocks, and edges represent control-flow transitions between basic blocks, typically induced by control-transfer instructions at the end of a block (e.g., jump and ret)~\cite{andriesse2016depth,stephens2016driller}.
Meanwhile, \name uses pseudo-code sequence as input to the LLM branch, as pseudo-code carries richer semantic information than raw assembly~\cite{li2025empiricalstudycodelarge}.
% }
By jointly representing assembly CFGs and pseudo-code, \name effectively captures both the structural and semantic information of binary patches, enabling accurate security patch identification.

\subsubsection{Graph Branch}
The \textit{\textbf{graph branch}} consists of three modules: a GNN module, an Adapter module, and a Cross-Attention~\cite{NIPS2017_3f5ee243} module.

(1) The \textit{GNN module} utilizes UniXcoder~\cite{guo2022unixcoder} to initialize the representation of each node (i.e., a basic block) in CFGs to generate the node embedding vector $h_i^{0}$, which can be calculated as $h_i^{0} = H^L(N_i)$ with $H^L(N_i)$ denoting the Pooler-layer output of UniXcoder.
Gated Graph Convolutional Network (GGCN)~\cite{bresson2017residual} is used to capture structural information in CFGs. Specifically, the node embeddings and edge embeddings are encoded by the GGCN, which is calculated as follows:
\begin{equation}\label{equ:gatedgcn}
\begin{gathered}
    h_i^{t+1} = ReLU(W_A^{t}h_i^{t} + \sum_{j \rightarrow i} \eta_{ij}^{t} \odot W_B^{t}h_j^{t}) \\
    e_{ij}^{t+1} = ReLU(W_U^{t} e_{ij}^{t} + W_V^{t} h_i^{t} + W_W^{t} h_j^{t}),\eta_{ij}^{t} = \sigma(e_{ij}^{t})
\end{gathered}
\end{equation}
Formula~\ref{equ:gatedgcn} represents the computation of GGCN, where at the $t+1$ layer, $h_i^{t+1}$ denotes the vector of node $i$, $e_{ij}^{t+1}$ denotes the edge vector from node $i$ to node $j$, and $W_A^{t}$, $W_B^{t}$, $W_U^{t}$, $W_V^{t}$, and $W_W^{t}$ are learnable weight parameters. $\eta_{ij}^{t}$ acts as the edge gate that can learn what edges are important for the graph learning. $ReLU$ denotes the Rectified Linear Unit (ReLU) activation function, $\sigma$ denotes the sigmoid activation function, and $\odot$ denotes the Hadamard point-wise multiplication. Finally, a pooling layer is employed over all node vectors of GGCN's final-layer $h^{T}$ to obtain the graph representation $f_g$.

(2) The \textbf{\textit{adapter module}} is the core mechanism of the graph branch, comprising a Query-Adapter, a Key-Adapter, a Value-Adapter, and a Gate Network.
The need for such adapters arises from the gap between the graph and LLM representations.
The graph representation $f_g$ is typically a dense embedding that encodes both graph topological structure and node-level semantics~\cite{kipf2016gcn}.
In contrast, the latent representations produced by LLMs empirically exhibit low intrinsic rank~\cite{hu2022lora,aghajanyan2020intrinsic}. 
Although $f_g$ remains highly information-expressive and thus has the potential to be mapped into the LLM's latent spaces, a simple linear transformation is generally insufficient to capture complex correspondences between two representation spaces of differing dimensionalities~\cite{glavavs2020non}.
Therefore, we employ Feed-Forward Neural Networks (FFNs)~\cite{rumelhart1986learning} as adapters to model complex nonlinear mappings via their universal approximation properties. 
Each adapter first projects the graph representation $f_g$ into a higher-dimensional space, applies a ReLU activation, and then maps it into LLM's latent spaces, with each adapter maintaining mutually independent parameters.
The three adapters respectively transform the graph representation $f_g$ to graph-derived hidden vectors $h_q'$, $h_k'$, and $h_v'$, each with a size equivalent to that of the LLM token vector, and then selectively activate them by a Gate Network (a Multilayer Perceptron). 
The gated vectors $h_k$ and $h_v$ are respectively concatenated to the LLM’s key $K_{LLM}$ and value $V_{LLM}$; vector $h_q$ is element-wise added to the LLM’s query $Q_{LLM}$ without altering the length of $Q_{LLM}$. 
The fused vectors $Q_{fusion}$, $K_{fusion}$, and $V_{fusion}$ subsequently participate in the LLM’s attention computation. 
Formula~\ref{equ:adapter} represents the computation of the adapters, where $Gate$ denotes the gate network, $FFN_q$, $FFN_k$, and $FFN_v$ are independent feed-forward networks.
\begin{equation}\label{equ:adapter}
\begin{gathered}
  g_g = Gate(f_g) \\
  h_{q,k,v} = FFN_{q,k,v}(f_g)\odot g_g
\end{gathered}
\end{equation}
Formula~\ref{equ:graph_llm_fusion} represents the computation of structure-guided joint representation, where $h_q \odot \mathbbm{1}_{q_{len}}$ represents broadcasting of $h_q$ to match the sequence length of $Q_{LLM}$, $\oplus$ is the element-wise addition, and $\parallel$ is the concatenation operation.
\begin{equation}\label{equ:graph_llm_fusion}
\begin{gathered}
   Q_{fusion} = h_q \odot \mathbbm{1}_{q_{len}} \oplus Q_{LLM} \\
  K_{fusion} = h_k \parallel K_{LLM}, ~
    V_{fusion} = h_v \parallel V_{LLM}
\end{gathered}
\end{equation}

This creative fusion design is grounded in the different functional roles of queries, keys, and values within the attention mechanism~\cite{NIPS2017_3f5ee243}. 
% \lqy{
Queries should maintain a one-to-one correspondence with the LLM’s input sequence length to ensure accurate querying for each input token, so the Query-Adapter injects controllable affine biases $h_q$ that shift the LLM's attention towards the structural tokens $h_k$ and $h_v$ without changing the query sequence length.
Keys and values act as an information pool for queries, so the Key and Value-Adapters concatenate the structural tokens $h_k$ and $h_v$ to the key/value sequences in attention to expand this pool with graph-specific knowledge, enabling the LLM to retrieve topological contexts (e.g., early exits) absent in the pseudo-code sequence. 
Furthermore, the Gate Network serves as a dynamic noise filter, suppressing irrelevant structural signals and effectively reducing false positives caused by subtle graph variations.
% }
% The three adapters are jointly optimized parameters within the entire graph branch, thus learning to map the graph representation $f_g$ to the LLM's latent spaces.

(3) The \textit{cross-attention module} computes cross-attention between the gated vectors $h_q$ and the hidden representations from the LLM’s final feed-forward layer $f_{llm}$, and finally employs a Multilayer Perceptron (MLP)~\cite{kruse2022multi} to produce logits for binary classification. Formula~\ref{equ:cross_attention} represents the computation of the prediction logits, where $CrossAttn$ denotes the cross-attention mechanism, $MLP$ denotes the MLP ~\cite{kruse2022multi}, and $\hat{y}$ denotes logits for binary classification.

\begin{equation}\label{equ:cross_attention}
\begin{aligned}
    \hat{y} &= \sigma( MLP(CrossAttn(h_q, f_{llm})) 
\end{aligned}
\end{equation}
% \begin{equation}\label{equ:cross_attention}
% \begin{aligned}
%     h_o &= CrossAttn(h_q, f_{llm}) \\
%     \hat{y} &= \sigma( MLP(h_o) )
% \end{aligned}
% \end{equation}

\subsubsection{LLM Branch}
The \textit{\textbf{LLM branch}} employs Qwen3-8B~\cite{yang2025qwen3} as the backbone to embed the instruction that contains the pre- and post-patch pseudo-code functions. 
% We select the Qwen3-8B because the Qwen3 family currently represents the state-of-the-art among open-source LLMs and achieves top performance on multiple code-related tasks~\cite{yang2025qwen3}.
This instruction queries whether the paired pseudo-code functions fix a vulnerability and requires the LLM to respond with ``yes'' or ``no''.
We argue that the semantic separation between ``yes'' and ``no'' in the LLM’s representation space is larger than that between the ``security'' and``non-security'', which leads to improved classification performance.

%%%% 需要说明下，为啥比直接concat图表征和序列表征的方法有效
Compared to directly concatenating graph and LLM representations, \name provides the LLM with structural information at a finer granularity (i.e., the token level). Through adapter-mediated alignment, the graph representation is mapped into the LLM’s latent space. Therefore, \name enables a tighter integration between structural and semantic representations.

% 训练算法
\subsection{Tow-stage Training} \label{section:two-stage}

Due to the significant disparity in parameter scale between the LLM and graph branches, the LLM branch attains strong performance with a smaller learning rate and fewer fine-tuning epochs. In contrast, graph branch requires a larger learning rate and more training epochs to reach optimal performance. This mismatch impedes the joint representation of the two branches. To mitigate this optimization imbalance, we adopt a two-stage training strategy that first fine-tunes the LLM branch and subsequently trains the graph branch with the LLM parameters frozen.

Concretely, we first perform instruction tuning~\cite{shengyu2023instruction} on the LLM branch, applying Low-Rank Adaptation (LoRA)~\cite{hu2022lora}. Given an instruction querying whether the pre- and post-patch pseudo-code functions constitute a security patch, the LLM is tasked with generating a short ``yes'' or ``no'' response. The parameters of the LLM branch are optimized by computing the Cross-Entropy loss between the generated token sequence and the target sequence (i.e., the ground-truth label), which enables the LLM to differentiate security from non-security patches.
Formula~\ref{equ:cross_entropy} represents the Cross-Entropy loss, where $N$ denotes the total number of training samples, $y_i$ denotes the target sequence for the sample $i$, $I_i$ denotes the input instruction to the sample $i$, and $\theta(\phi_\textit{l})$ refers to the parameters of the LLM.
\begin{equation}\label{equ:cross_entropy}
   \mathcal{L}_{\textit{ce}}(\phi_\textit{l}) = - \frac{1}{N} \sum_{i=1}^{N} \log P(y_i \mid I_i; \theta(\phi_\textit{l}))
\end{equation}
After fine-tuning the LLM branch, we freeze its parameters and proceed to train the graph branch. The parameters of the graph branch are optimized by calculating the Binary Cross-Entropy (BCEWithLogits) loss between the predictions of the \name model and the ground-truth labels. 
Formula~\ref{equ:bcewithlogits} represents the BCEWithLogits loss, where $y_i$ denotes the ground-truth label (either 0 or 1) for the sample $i$, the joint input denoted as $\langle G_i, I_i \rangle$, consists of the pre- and post-patch graphs along with the instruction, $\sigma(z_i)$ represents the prediction of the \name model, and $\phi_{\textit{g}}$ refers to the parameters of the graph branch.
% \begin{equation}\label{equ:bcewithlogits}
% \begin{aligned}
%     \mathcal{L}_{\textit{bce}}(\phi_{\textit{g}}) = - \frac{1}{N} \sum_{i=1}^{N} \left[ y_i \log \sigma(z_i) + (1 - y_i) \log (1 - \sigma(z_i)) \right], ~
%     \sigma(z_i) = P(y_i \mid \langle G_i, I_i \rangle; \phi_{\textit{g}})
% \end{aligned}
% \end{equation}
\begin{equation}\label{equ:bcewithlogits}
\begin{gathered}
  \sigma(z_i) = P(y_i \mid \langle G_i, I_i \rangle; \phi_{\textit{g}}) \\
  \mathcal{L}_{\textit{bce}}(\phi_{\textit{g}}) = - \frac{1}{N} \sum_{i=1}^{N} \left[ y_i \log \sigma(z_i) + (1 - y_i) \log (1 - \sigma(z_i)) \right]
\end{gathered}
\end{equation}

% Formula~\ref{equ:lora} represents the LoRA parameterization used to update the LLM, where $\theta_0$ denotes the pre-trained parameters of the LLM, and $\Delta(\phi)$ corresponds to the learnable update. In this parameterization, $\Delta(\phi)$ is decomposed into two low-rank matrices, $A$ and $B$, with rank $r$ and scaled by a factor of $\alpha/r$, enabling an efficient parameter update.
% \begin{equation}
%     \theta(\phi)=\theta_0 + \Delta(\phi),\qquad
%     \Delta(\phi) = \frac{\alpha}{r}\, A B,
%     \label{equ:lora}
% \end{equation}

\section{Experimental Setup} \label{section:setup}

\subsection{Research Questions}

In this section, we investigate the effectiveness of \name by comparing it with the state-of-the-art binary SPD methods and focus on answering the following research questions (RQs).

\begin{itemize}

% 对比实验（主试验）
\item \textbf{RQ1}: How effective is \name compared with existing binary SPD methods?

% 新增
% \lqy{
\item \textbf{RQ2}: How efficient is \name compared with existing binary SPD methods in terms of time and cost?
% }

% 分析实验，分析在各种漏洞对应的安全补丁上的识别效果
\item \textbf{RQ3}: How effective is \name compared with existing binary SPD methods in identifying patches corresponding to different vulnerability types?

% 泛化性实验
\item \textbf{RQ4}: How effective is \name in adapting existing LLMs for binary SPD?

% 消融实验，消融各个模块
\item \textbf{RQ5}: What is the contribution of each design of \name?

\end{itemize}

\subsection{Baselines}

% LLMs
\textbf{\textit{LLM-based Methods}}: 
Previous research~\cite{li2025empiricalstudycodelarge} has demonstrated that code LLMs achieve excellent performance on binary SPD tasks through fine-tuning. 
We therefore select six code LLMs with varying scales that have exhibited strong performance on binary SPD tasks, including Qwen2.5-Coder-0.5B-Instruct~\cite{qwen2.5coder}, Qwen2.5-Coder-7B-Instruct~\cite{qwen2.5coder}, LLM-Compiler-7B~\cite{cummins2025llm}, LLM-Compiler-7B-ftd~\cite{cummins2025llm}, Yi-Coder-9B-Chat~\cite{yi-coder}, and LLM4Decompile-9B-v2~\cite{tan2024llm4decompile}. 
To broaden the comparison, we also evaluate two powerful closed-source foundation models, GPT-4o~\cite{achiam2023gpt} and GPT-5-mini~\cite{gpt5} via zero-shot prompting.
% We include Qwen3-8B~\cite{yang2025qwen3} as a baseline, since it is currently the state-of-the-art open-source LLM.
% \begin{itemize}
% \item \textbf{Qwen2.5-Coder-0.5B-Instruct} and \textbf{Qwen2.5-Coder-7B-Instruct}~\cite{qwen2.5coder} build on Qwen2.5 and are trained with a larger volume of code tokens, which significantly improves code generation and reasoning. 
% \item \textbf{Qwen3-8B}~\cite{yang2025qwen3} is the latest generation of LLMs in the Qwen series, offering a comprehensive suite of dense and Mixture-of-Experts (MoE)~\cite{zhou2022mixture} models.
% \item \textbf{Yi-Coder-9B-Chat}~\cite{yi-coder} supports 52 major programming languages, including C, C++, and assembly.
% \item \textbf{LLM4Decompile-9B-v2}~\cite{tan2024llm4decompile} is a state-of-the-art binary code model that targets decompilation from assembly to C source.
% \item \textbf{LLM-Compiler-7B} and \textbf{LLM-Compiler-7B-ftd}~\cite{cummins2025llm} are state-of-the-art binary code models built on CodeLlama~\cite{codellama} with improved capability for assembly code, optimization, and compiler reasoning.
% \end{itemize}

% BinGo
\textbf{\textit{Graph-based Methods}}: We select BinGo~\cite{he2024bingo} as a baseline, as it exhibits excellent performance on binary SPD tasks and demonstrates good robustness across multiple optimization levels. 
% We briefly introduce BinGo.
% \begin{itemize}
% \item \textbf{BinGo}~\cite{he2024bingo} is a graph-based security patch detection system designed for binary code. 
% % It represents binary code as Code Property Graphs (CPGs)~\cite{he2024bingo} that integrate control flow, control dependency, and data dependency graphs to capture program semantics comprehensively. 
% It identifies patch-related code segments between pre-patch and post-patch binaries by leveraging graph representation learning, using graph neural networks (GNNs)~\cite{ggnn} to analyze Code Property Graphs (CPGs), and comparing the structural and semantic changes between different binary versions.
% \end{itemize}

% PatchRNN & LLMDA
\textbf{\textit{Modified Source-code Methods}}: Since rule-based method 1dvul~\cite{peng20191dvul} and pattern-based method SPAIN~\cite{xu2017spain} are not publicly available. To broaden the comparison beyond binary SPD, we select two source-code SPD methods, PatchRNN~\cite{wang2021patchrnn} and LLMDA~\cite{tang2023just}, which can be adapted to the binary SPD task.
% \begin{itemize}
% \item \textbf{PatchRNN}~\cite{wang2021patchrnn} encodes reconstructed diff code with twin Recurrent Neural Networks (RNNs)~\cite{sherstinsky2020fundamentals} and encodes commit messages with a TextRNN. The model concatenates these embeddings and uses a classifier to predict the patch's security relevance.
% \item \textbf{LLMDA}~\cite{tang2023just} leverages LLMs both to generate natural-language explanations of code changes for data augmentation and to encode code changes and commit messages via CodeT5+~\cite{wang2023codet5+} and LLaMa-7b~\cite{touvron2023llama}. It aligns these multimodal embeddings with a PT-Former module, injects a label-wise instruction token, and trains the classifier using stochastic batch contrastive learning (SBCL). It yields stronger semantic understanding and achieves state-of-the-art results on source-code SPD.
% \end{itemize}

\subsection{Evaluation Metrics}
We choose the following metrics to evaluate the performance of all binary SPD methods.

\begin{itemize}
    \item \textbf{Accuracy}: $Acc=\frac{TP+TN}{TP+TN+FN+FP}$. This metric measures the proportion of correctly classified samples among all evaluated samples, where $TP$ and $TN$ represent the counts of true positive and true negative samples, and $TP+TN+FN+FP$ represents the total number of samples.

    \item \textbf{Precision}: $Pre=\frac{TP}{TP+FP}$. This metric measures the correctness of predictions in positive samples, where $TP+FP$ represents the total number of samples predicted as positive.

    \item \textbf{Recall}: $Rec=\frac{TP}{TP+FN}$. This metric measures the proportion of actual positive samples that are correctly identified, where $FN$ represents the count of false negative samples.

    \item \textbf{F1 score}: $F1=\frac{2\times Precision\times Recall}{Precision+Recall}$. This metric measures the comprehensive performance of positive-class predictions by acting as the harmonic mean of precision and recall.
     
    \item \textbf{False Positive Rate (FP Rate)}: $FPR=\frac{FP}{FP+TN}$. This metric measures the ratio of erroneously classified positive samples, where $FP$ represents the count of false positive samples.
\end{itemize}

\subsection{Dataset Selection}

\textbf{\textit{Dataset for Training}}: We use the dataset from \citet{li2025empiricalstudycodelarge} as the training and validation sets for \name and the baselines. This dataset contains 16,957 training samples and 1,248 validation samples, with a positive-to-negative sample ratio of 1:1.3. We use the training set to train the models and the validation set to select the models with the best performance.

\textbf{\textit{Benchmark for Evaluation}}: After completing the training process, we evaluate \name and the baselines on the benchmark constructed in Section~\ref{section:bench}, as this benchmark can far more effectively reflect the model performance in real-world scenarios.

\subsection{Implementation Details} 
% lqy: 我认为动作应该和实验结果完全独立，因此这一节写的详细一些

%%%% RQ1
To answer \textbf{RQ1}, we compare \name with three categories of binary SPD methods, including LLM-based, graph-based, and modified source-code methods. 
To evaluate the LLM-based baselines, we adopt the hyperparameters provided by \citet{li2025empiricalstudycodelarge} and utilize the pseudo-code representation data as input to LLMs, as they have demonstrated that pseudo-code is better suited for code LLMs on binary SPD tasks.
To evaluate the graph-based and modified source-code baselines, we use their publicly available source code, following the papers that use the assembly-code CPG as input to BinGo, the pseudo-code diff as input to PatchRNN, and LLMDA. In closed-source scenarios, commit messages are unavailable, so we omit them when evaluating the modified source-code baselines. 
To evaluate \name, the assembly-code CFGs are used as input for the graph branch, and the instruction composed of pseudo-code functions is used as input for the LLM branch. 
The hyperparameters used for \name to training are as follows: 
The fine-tuning epoch for the LLM is set to $3$, while that for the graph branch is $8$. 
The learning rate is configured as $1 \times 10^{-4}$ for the LLM and $5 \times 10^{-5}$ for the graph branch. 
The batch size is set to $4$ for both branches.

%%%% RQ2新增
% \lqy{
To answer \textbf{RQ2}, for methods that require training, we report the time cost of both training and inference for each method, and convert GPU usage into monetary cost based on AutoDL’s pricing (4×A100 40GB: \$1.85 per hour,~\url{https://www.autodl.com}). For methods that do not require training, we directly report the OpenAI API call runtime and monetary cost.
% \footnote{\url{https://openai.com/api}}}

%%%% RQ3
To answer \textbf{RQ3}, we extract the Common Weakness Enumerations (CWEs)~\cite{cwe} associated with every positive sample in our benchmark and assign each CWE to its corresponding category. The benchmark spans 11 vulnerability categories and comprises 26 distinct CWE types. 
We subsequently evaluate the identification performance of \name against the three best-performing baselines on security patches grouped by corresponding vulnerability type. Since precision more accurately reflects a model’s ability to identify security patches, we select the best-performing baselines based on this metric.

%%%% RQ4
To answer \textbf{RQ4}, we select ten widely used open-source LLMs drawn from four very popular model families, including Llama~\cite{touvron2023llama}, Qwen~\cite{qwen2}, DeepSeek~\cite{deepseekllm}, and Yi~\cite{yi-coder}.
The selected models are Llama-3.2-3B-Instruct, Llama-3.1-8B-Instruct, LLM-Compiler-7B, Qwen2.5-Coder-0.5B-Instruct, Qwen2.5-Coder-7B-Instruct, Qwen3-8B, DeepSeek-Coder-1.3B-Instruct, DeepSeek-Coder-7B-Instruct-v1.5, Yi-Coder-9B-Chat, and LLM4Decompile-9B-v2. 
In particular, LLM-Compiler-7B and LLM4Decompile-9B-v2 are derived from CodeLlama-7B and Yi-Coder-9B-Chat, respectively, and have been pre-fine-tuned for binary-related tasks.
We employ each model as the backbone of the LLM branch to evaluate variants of \name and investigate whether the \name framework generalizes to existing open-source foundation language models.

%%%% RQ5
To answer \textbf{RQ5}, we conduct a comprehensive ablation study by 
excluding the joint representation of assembly code and pseudo-code (i.e., w/o Graph branch and w/o LLM branch), 
% \lqy{
replacing adaptive fusion with naive concatenation (i.e., w/o Adaptive fusion),
% }
removing individual modules in the graph branch (i.e., w/o Adapters, w/o Gate, and w/o Cross-attention), 
and disabling the two-stage training strategy (i.e., w/o Two-stage training). 

Note that when excluding the graph branch (w/o Graph branch), the LLM takes as input the serialized assembly code followed by the pseudo-code, which corresponds to the straightforward idea discussed in Section~\ref{section:motivation}; 
% \lqy{
and when replacing adaptive fusion with naive concatenation (w/o Adaptive fusion), we remove all fusion components in the graph branch, including three adapters, the gate network, and the cross-attention network, and directly concatenate the GGCN graph representation with the LLM's last hidden state to form the joint representation for prediction. 
% }

%%%% Experiment environments
All experiments are conducted on a server running Ubuntu OS, equipped with an Intel Xeon Platinum 8260 CPU and four NVIDIA A100 Tensor Core GPUs, each with 40GB of memory and CUDA version 12.4.

\section{Experimental Results} \label{section:results}

%%%% 主试验
\subsection{RQ1: \name vs. Baselines in Effectiveness}

% The experimental results of the comparison between \name and the three categories of baselines are shown in Table~\ref{tab:rq1}.
Table~\ref{tab:rq1} presents the comparison results of \name against the three categories of baselines.
\begin{table*}[htbp]
\centering
\caption{Experimental results of \name and the baselines}
\label{tab:rq1}
\resizebox{0.65
\textwidth}{!}{%
\begin{tabular}{cccccc}
\toprule
\textbf{Method} & \textbf{Accuracy$\uparrow$} & \textbf{Precision$\uparrow$} & \textbf{Recall$\uparrow$} & \textbf{F1 Score$\uparrow$} & \textbf{FP Rate$\downarrow$} \\
\cmidrule(lr){1-6}\morecmidrules\cmidrule(lr){1-6}

Qwen2.5-Coder-0.5B-Instruct & 0.528 & 0.597 & 0.603 & 0.600 & 0.579 \\
Qwen2.5-Coder-7B-Instruct & 0.617 & 0.658 & 0.726 & 0.690 & 0.537 \\
LLM-Compiler-7B & 0.559 & 0.614 & 0.672 & 0.642 & 0.601 \\
LLM-Compiler-7B-ftd & 0.477 & 0.553 & 0.576 & 0.564 & 0.663 \\
% Qwen3-8B & 0.643 & 0.685 & 0.726 & 0.705 & 0.475 \\
Yi-Coder-9B-Chat & 0.758 & 0.733 & 0.924 & 0.818 & 0.477 \\
LLM4Decompile-9B-v2 & 0.542 & 0.644 & 0.491 & 0.557 & 0.348 \\
GPT-4o & 0.592 & 0.612 & 0.833 & 0.706 & 0.751 \\
GPT-5-mini & 0.598 & 0.623 & 0.800 & 0.700 & 0.689 \\
\midrule

BinGo & 0.610 & 0.601 & \textbf{1.0} & 0.751 & 0.944 \\
\midrule

PatchRNN & 0.553 & 0.628 & 0.587 & 0.607 & 0.537 \\
LLMDA & 0.491 & 0.569 & 0.547 & 0.558 & 0.589 \\
\midrule

\textbf{\name} & \textbf{0.854} & \textbf{0.823} & 0.957 & \textbf{0.885} & \textbf{0.293} \\

\bottomrule
\end{tabular}
}
% \begin{tablenotes}
%     \scriptsize
%     \raggedright
%     \item The best baseline is colored with \colorbox{gray!40}{gray}.
%     \par
% \end{tablenotes}
\end{table*}

\textbf{\textit{Overall Results}}: 
% lqy2ycc: overall results中可以只说我们方法自己的性能，先不做比较，因为后面会展开比较。
% ycc: understand
\name demonstrates strong effectiveness on the cross-project, cross-domain benchmark, consistently outperforming all baselines and achieving an accuracy of 0.854, an F1 score of 0.885, and a false positive rate of 0.293.

\textbf{\textit{\name vs. LLM-based Methods}}: 
Although LLMs are the strongest baselines, \name consistently outperforms LLM-based methods that are fine-tuned exclusively on pseudo-code datasets across all evaluation metrics.
Specifically, \name improves accuracy by 38.41\% and 52.77\%, and F1 score by 28.26\% and 37.85\%, relative to Qwen2.5-Coder-7B-Instruct and LLM-Compiler-7B, respectively.
Notably, compared with the best-performing baseline Yi-Coder-9B-Chat, \name outperforms it by 12.66\% in accuracy and 8.19\% in F1 score, accompanied by a 38.57\% reduction in the false positive rate.
Furthermore, compared with state-of-the-art foundation language models, \name consistently outperforms them across all metrics, particularly achieving improvements of 42.81\% in accuracy and 57.47\% in false positive rate over the best-performing GPT-5-mini.
These results suggest that solely fine-tuning LLMs on pseudo-code datasets does not guarantee effectiveness across all LLMs.

\textbf{\textit{\name vs. Graph-based Methods}}: 
Compared with BinGo, \name outperforms it by 40.00\% in accuracy and 36.94\% in precision. Although Bingo attains a recall of 1.0, it also has a false-positive rate of 0.944.
By contrast, our method not only maintains very high recall but also achieves a substantially lower false positive rate. 
These results suggest that relying solely on graph networks to exploit structural information in assembly code is insufficient for achieving strong performance in binary SPD.

\textbf{\textit{\name vs. Modified Source-code Methods}}: 
Among all types of baselines, \name achieves the greatest performance gain over the modified source-code methods.
Specifically, compared with PatchRNN, \name improves accuracy by 54.43\%, increases F1 score by 45.80\%, and reduces the false positive rate by 45.44\%. 
These results indicate that methods designed for source-code SPD do not generalize well to binary SPD tasks.

\begin{figure}[t!]
  \centering
  \includegraphics[width=0.9\linewidth]{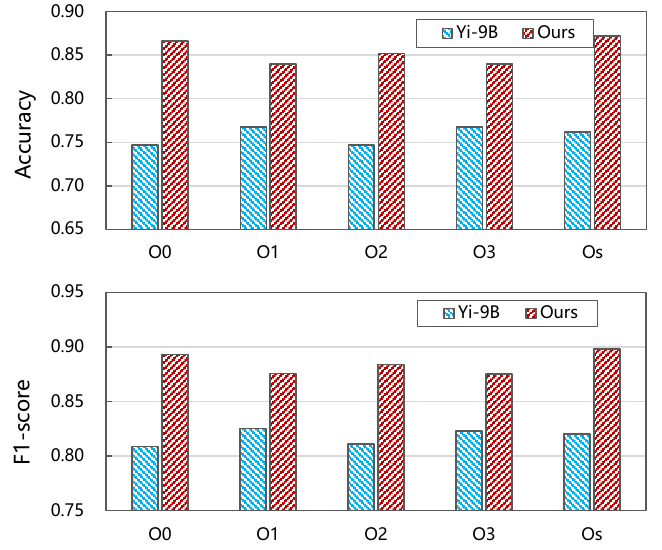}
  \vspace{-0.5em}
  \caption{Accuracy and F1 score of \name compared with the best baseline across different optimization levels.}
  \label{fig:optim_acc_f1}
  \vspace{-0.5em}
\end{figure}

% \lqy{
\textbf{\textit{\name vs. Best Baseline across Optimization Levels:}}
Additionally, Figure~\ref{fig:optim_acc_f1} presents the accuracy and F1 score of \name compared with the best-performing baseline across different optimization levels, achieving improvements in both metrics.
Although \name’s performance drops as the optimization level increases, it still delivers clear improvements over the strongest baseline.
To investigate this trend, we compute assembly-code CFG features and pseudo-code lengths for each optimization level.
As shown in Table~\ref{tab:opt_stats}, both the complexity of the assembly-code CFG and the pseudo-code length increase with higher optimization levels. These observations help explain why \name performs better on O0 and Os than on O1, O2, and O3.
% }
\begin{table}[t!]
\centering
\caption{CFG and token statistics across optimization levels}
\label{tab:opt_stats}
\resizebox{0.4\textwidth}{!}{
\begin{tabular}{cccc}
\toprule
\textbf{Optim. Level} & \textbf{Avg. Node} & \textbf{Avg. Edge} & \textbf{Avg. Token} \\
\cmidrule{1-4}\morecmidrules\cmidrule{1-4}

O0 & 95.7 & 147.6 & 4049.0 \\
O1 & 99.1 & 154.1 & 4385.7 \\
O2 & 99.4 & 154.5 & 4519.5 \\
O3 & 101.7 & 158.3 & 4628.2 \\
Os & 98.2 & 151.8 & 4355.2 \\

\bottomrule
\end{tabular}
}
\end{table}

% \begin{figure*}[htbp]
%   \centering
%   % 两侧留白更大：总宽度小于 \textwidth
%   \begin{minipage}[t]{0.4\textwidth}
%     \centering
%     \includegraphics[width=\linewidth]{figures/2_Figure2_OursVsYi_Acc.pdf}
%   \end{minipage}\hspace{0.02\textwidth}
%   \begin{minipage}[t]{0.4\textwidth}
%     \centering
%     \includegraphics[width=\linewidth]{figures/3_Figure3_OursVsYi_F1.pdf}
%   \end{minipage}
%   \caption{Accuracy and F1 score of \name compared with the best baseline across different optimization levels.}
%   \label{fig:optim_acc_f1}
% \end{figure*}

\begin{tcolorbox}[colback=background!50,
        colframe=edge,
        width=\columnwidth,% total width
        boxrule = 0.3mm,
        top = 3pt, bottom=3pt, left=3pt, right=3pt
    ]
    \textbf{Answer to RQ1:} 
    \name exhibits effectiveness on binary SPD tasks, achieving the best performance compared against existing LLM-based, graph-based, and modified source-code binary security patch detection methods. 
\end{tcolorbox}

%%%% 新增
% \lqy{
\subsection{RQ2: \name vs. Baselines in Efficiency and Cost}
Table~\ref{tab:rq1-2} reports the end-to-end training time, average per-sample inference latency, and total monetary cost (training plus inference/API calls) for \name and the three categories of baselines. Overall, \name attains a practical inference latency of 0.67s/sample, which is in the same range as the baselines (0.01 to 0.29s/sample). Its training time (11h22m) is moderately higher than comparably sized open-source LLM baselines (e.g., 7B to 9B models), which we attribute to our two-stage training procedure. Despite this additional training overhead, \name remains cost-effective overall (\$21.62) and is substantially cheaper than API-based solutions such as GPT-4o (\$74.00), indicating a favorable efficiency cost trade-off.

% }
\begin{tcolorbox}[colback=background!50,
        colframe=edge,
        width=\columnwidth,% total width
        boxrule = 0.3mm,
        top = 3pt, bottom=3pt, left=3pt, right=3pt
    ]
   \textbf{Answer to RQ2:} 
   \name is practical in both efficiency and cost for binary SPD, maintaining acceptable inference latency and remaining overall cost-effective, especially compared with API-based LLM solutions. 
\end{tcolorbox}

%%%% 分析实验
\subsection{RQ3: \name vs. Baselines across Types of Patches}

% The experimental results of the comparison between \name and the baselines in identifying patches corresponding to different vulnerability types are shown in Table~\ref{tab:rq2}.
Table~\ref{tab:rq2} presents the accuracy of \name and the top-3 baselines in identifying patches across different vulnerability types.
\begin{table}[t!]
\centering
\caption{Time and Cost Comparison of \name and Baselines}
\label{tab:rq1-2}
\resizebox{0.48\textwidth}{!}{
\begin{tabular}{cccc}
\toprule
\textbf{Method} & \textbf{Training Time} & \textbf{Avg. Infer. Time}  & \textbf{Cost}\\
\cmidrule(lr){1-4}\morecmidrules\cmidrule(lr){1-4}
Qwen0.5B & 59m3s & 0.03s & 1.88 \\
Qwen7B & 4h56m & 0.20s & 9.30 \\
LLMC7B & 10h21m & 0.22s & 19.35 \\
LLMC7B-ftd & 5h33m & 0.22s & 10.47 \\
Yi9B & 7h35m & 0.29s & 14.28 \\
LLM4D9B & 7h36m & 0.29s & 14.31 \\
GPT-4o & - & 0.15s & 74.00 \\
GPT-5-mini & - & 0.18s & 22.00 \\
\midrule
BinGo & 25m59s & 0.21s & 0.99 \\
\midrule
PatchRNN & 12m36s & 0.01s & 0.39 \\
LLMDA & 3h40m & 0.08s & 6.85 \\
\midrule
\textbf{\name} & 11h22m & 0.67s & 21.62 \\
\bottomrule
\end{tabular}
}
\end{table}

\begin{table*}[htbp]
\centering
\caption{Accuracy of \name and the baselines over different vulnerability type patches}
\label{tab:rq2}
\resizebox{0.8\textwidth}{!}{%
\begin{tabular}{ccccccc}
\toprule
\textbf{Vulnerability Type} & \textbf{CWE ID} & \textbf{Acc.} & \textbf{\name} & \textbf{Yi9B} & \textbf{Qwen7B} & \textbf{LLM4D9B} \\
\cmidrule(lr){1-7}\morecmidrules\cmidrule(lr){1-7}

Buffer Overflow & 119, 120, 122, 125, 787 & 67.5\% & 0.935 & 0.885 & 0.654 & 0.523\\
Resource Management Errors & 399, 400, 401, 770, 772, 908 & 14.3\% & 1.0 & 0.974 & 0.833 & 0.442\\
Control-flow Errors & 674, 834, 835 & 3.9\% & 1.0 & 1.0 & 0.708 & 0.569\\
Improper Input Validation & 20 & 3.9\% & 0.985 & 1.0 & 0.892 & 0.523\\
Numberic Errors & 190, 193, 369 & 3.9\% & 1.0 & 0.815 & 0.769 & 0.092\\
Invalid Memory Free & 415, 416 & 2.1\% & 1.0 & 0.971 & 0.829 & 0.457\\
Runtime Check Errors & 252, 617 & 1.8\% & 0.967 & 1.0 & 0.867 & 0.433\\
Information Leak & 200 & 1.2\% & 1.0 & 1.0 & 0.75 & 0.35\\
Null Pointer Dereference & 476 & 0.9\% & 1.0 & 1.0 & 0.733 & 0.4\\
Improper Access Control & 284 & 0.3\% & 1.0 & 1.0 & 0.6 & 0.8\\
Type Errors & 704 & 0.3\% & 1.0 & 1.0 & 1.0 & 0.0\\

\bottomrule
\end{tabular}
}
\end{table*}

Overall, \name outperforms the top-3 baselines across nine patch types.
For the two most frequent patch types, Buffer Overflow and Resource Management Errors, \name exceeds the top-3 baselines by 5.65\%, 42.97\%, and 78.78\%; and by 2.67\%, 20.05\%, and 126.24\%, respectively.
We observe that: (1) \name achieves highly accurate identification for both frequent patch types (e.g., Buffer Overflow) and less frequent ones (e.g., Type Errors), indicating that the joint representation of structural and semantic information enables \name to learn the patterns of security patches effectively.
(2) The only two patch types where \name underperforms the top-1 baseline are runtime check errors and improper input validation. 
Patches of these types often do not manifest as compile-time faults: runtime check errors in particular are triggered only under specific dynamic inputs, while improper input validation typically does not exhibit overt errors, instead requiring multiple repository-level dependencies understanding. 
Thus, both types require precise semantic understanding; however, structural information in low-level code may interfere with the LLM’s ability to interpret such semantics

\begin{tcolorbox}[colback=background!50,
        colframe=edge,
        width=\columnwidth,% total width
        boxrule = 0.3mm,
        top = 3pt, bottom=3pt, left=3pt, right=3pt
    ]
    \textbf{Answer to RQ3:} 
    \name outperforms the top baselines in identifying security patches, especially with a higher number of security patches. The missed security patches are usually undetectable at compile time and demand runtime-aware program behaviour understanding. In such cases, introducing low-level code structural information may hinder the LLM’s ability to capture those high-level semantics.
\end{tcolorbox}

%%%% 泛化性实验
\subsection{RQ4: Effectiveness of \name in Adapting LLMs} \label{section:rq3}

% The experimental results of the \name employing different models as the backbone of the LLM branch are shown in Table~\ref{tab:rq3}.
Table~\ref{tab:rq3} presents the performance of \name with different models as the backbone of the LLM branch.
\begin{table*}[htbp]
\centering
\caption{Experimental results of \name in adapting different LLMs}
\label{tab:rq3}
\resizebox{0.8\textwidth}{!}{%
\begin{tabular}{ccccccc}
\toprule
\textbf{Family} & \textbf{Variant} & \textbf{Accuracy$\uparrow$} & \textbf{Precision$\uparrow$} & \textbf{Recall$\uparrow$} & \textbf{F1 Score$\uparrow$} & \textbf{FP Rate$\downarrow$} \\
\cmidrule(lr){1-7}\morecmidrules\cmidrule(lr){1-7}

\multirow{3}{*}{Yi} 

& \textit{w/ Yi-Coder-1.5B-Chat} & 0.64 \footnotesize (\textcolor{green}{+28.0\%}) & 0.69 \footnotesize (\textcolor{green}{+16.9\%}) & 0.68 \footnotesize (\textcolor{green}{+\textbf{44.7}\%}) & 0.69 \footnotesize (\textcolor{green}{+\textbf{30.2}\%}) & 0.43 \footnotesize (\textcolor{green}{+8.5\%}) \\ 
& \textit{w/ Yi-Coder-9B-Chat} & 0.66 \footnotesize (\textcolor{red}{-13.2\%}) & 0.70 \footnotesize (\textcolor{red}{-5.2\%}) & 0.74 \footnotesize (\textcolor{red}{-19.5\%}) & 0.72 \footnotesize (\textcolor{red}{-12.1\%}) & 0.46 \footnotesize (\textcolor{green}{+2.9\%}) \\ 
& \textit{w/ LLM4Decompile-9B-v2} & 0.52 \footnotesize (\textcolor{red}{-3.5\%}) & 0.63 \footnotesize (\textcolor{red}{-3.0\%}) & 0.47 \footnotesize (\textcolor{red}{-4.5\%}) & 0.54 \footnotesize (\textcolor{red}{-3.8\%}) & 0.40 \footnotesize (\textcolor{red}{-14.7\%}) \\

\midrule
\multirow{2}{*}{DeepSeek} 
& \textit{w/ DS-Coder-1.3B-I} & 0.57 \footnotesize (\textcolor{green}{+9.1\%}) & 0.63 \footnotesize (\textcolor{green}{+5.7\%}) & 0.66 \footnotesize (\textcolor{green}{+12.1\%}) & 0.65 \footnotesize (\textcolor{green}{+9.0\%}) & 0.55 \footnotesize (\textcolor{green}{+2.9\%}) \\
& \textit{w/ DS-Coder-7B-I-v1.5} & 0.65 \footnotesize (\textcolor{green}{+22.7\%}) & 0.72 \footnotesize (\textcolor{green}{+16.1\%}) & 0.66 \footnotesize (\textcolor{green}{+28.7\%}) & 0.68 \footnotesize (\textcolor{green}{+22.6\%}) & 0.37 \footnotesize (\textcolor{green}{+6.6\%}) \\
& \textit{w/ DS-Coder-V2-Lite-I} & 0.69 \footnotesize (\textcolor{green}{+2.2\%}) & 0.70 \footnotesize (\textcolor{green}{+2.4\%}) & 0.83 \footnotesize (\textcolor{red}{-1.1\%}) & 0.76 \footnotesize (\textcolor{green}{+1.0\%}) & 0.50 \footnotesize (\textcolor{green}{+6.8\%}) \\

\midrule
\multirow{3}{*}{Llama} 
& \textit{w/ Llama-3.2-3B-I} & 0.60 \footnotesize (\textcolor{green}{+6.6\%}) & 0.66 \footnotesize (\textcolor{green}{+6.9\%}) & 0.66 \footnotesize (\textcolor{red}{-2.5\%}) & 0.66 \footnotesize (\textcolor{green}{+2.2\%}) & 0.48 \footnotesize (\textcolor{green}{+30.4\%}) \\
& \textit{w/ LLM-Compiler-7B} & 0.61 \footnotesize (\textcolor{green}{+9.3\%}) & 0.66 \footnotesize (\textcolor{green}{+8.0\%}) & 0.69 \footnotesize (\textcolor{green}{+2.4\%}) & 0.68 \footnotesize (\textcolor{green}{+5.2\%}) & 0.50 \footnotesize (\textcolor{green}{+25.5\%}) \\
& \textit{w/ Llama-3.1-8B-I} & 0.61 \footnotesize (\textcolor{red}{-3.2\%}) & 0.65 \footnotesize (\textcolor{red}{-0.8\%}) & 0.71 \footnotesize (\textcolor{red}{-6.6\%}) & 0.68 \footnotesize (\textcolor{red}{-3.6\%}) & 0.54 \footnotesize (\textcolor{green}{+4.7\%}) \\

\midrule
\multirow{3}{*}{Qwen} 
& \textit{w/ Qwen2.5-Coder-0.5B-I} & 0.59 \footnotesize (\textcolor{green}{+11.0\%}) & 0.64 \footnotesize (\textcolor{green}{+6.9\%}) & 0.68 \footnotesize (\textcolor{green}{+13.0\%}) & 0.66 \footnotesize (\textcolor{green}{+9.8\%}) & 0.55 \footnotesize (\textcolor{green}{+5.8\%}) \\
& \textit{w/ Qwen2.5-Coder-7B-I} & 0.75 \footnotesize (\textcolor{green}{+20.8\%}) & 0.74 \footnotesize (\textcolor{green}{+12.2\%}) & 0.88 \footnotesize (\textcolor{green}{+20.8\%}) & 0.80 \footnotesize (\textcolor{green}{+16.1\%}) & 0.44 \footnotesize (\textcolor{green}{+17.3\%}) \\
& \textit{\textbf{w/ Qwen3-8B}} & \textbf{0.85} \footnotesize (\textcolor{green}{+\textbf{32.8}\%}) & \textbf{0.82} \footnotesize (\textcolor{green}{+\textbf{20.2}\%}) & \textbf{0.96} \footnotesize (\textcolor{green}{+31.8\%}) & \textbf{0.89} \footnotesize (\textcolor{green}{+25.5\%}) & \textbf{0.29} \footnotesize (\textcolor{green}{+\textbf{38.3}\%}) \\

\bottomrule
\end{tabular}
}
\end{table*}

Experimental results show that \name is compatible with many LLMs, achieving noticeable enhancements when employing various models from the Llama, Qwen, and DeepSeek families as the backbone of the LLM branch.
Notably, \name with Qwen3-8B (w/ Qwen3-8B) achieves the best results, improving accuracy, F1 score, and false positive rate by 32.8\%, 25.5\%, and 38.3\%, respectively. Additionally, Qwen2.5-Coder-7B-Instruct from the Qwen family also delivers considerable gains, with improvements of 20.8\%, 16.1\%, and 17.3\% across these three metrics.
Unexpectedly, \name with the best baseline Yi-Coder-9B-Chat (w/ Yi-Coder-9B-Chat) does not achieve further performance gains, as all other metrics decrease except for a slight improvement in the false positive rate. In contrast, Yi-Coder-1.5B-Chat from the same Yi family shows substantial improvements, achieving the largest gains among all LLMs in recall and F1 score, with increases of 44.7\% and 30.2\%, respectively. 

We infer that the adaptation of \name may depend on the model family of its LLM branch.
Although most modern LLMs adopt the standard Transformer architecture and follow broadly standardized data collection and training pipelines, vendor-specific design, such as architecture (e.g., Mixture-of-Experts) and pretraining schedules (e.g., domain-adaptive pretraining), can induce measurable differences in the hidden representation spaces of LLMs~\cite{klabunde2023towards,dai2024representational}.
\name's graph branch is designed to supply structural information to the LLM, but UniXcoder, used to encode node semantics, also inhabits its own representation space. UniXcoder's representations may not align well with those of certain target LLMs, thereby limiting the effectiveness of the joint representation.
%\lqy{
We further compare the Centered Kernel Alignment (CKA) similarity~\cite{kornblith2019similarity} between UniXcoder embeddings and those of the Yi family models, as well as with Qwen3-8B.
The results show that UniXcoder aligns substantially worse with the Yi-family models than with Qwen3-8B: the similarities between UniXcoder and Yi-Coder-1.5B-Chat, Yi-Coder-1.5B-Chat, and LLM4Decompile-9B-v2 are 0.16, 0.17, and 0.14, respectively, whereas the similarity with Qwen3-8B is 0.22, which is markedly higher than those of the Yi-family models.
This finding suggests that a mismatch between the graph-branch embedding model and the target LLM can degrade \name’s performance.
% }

\begin{tcolorbox}[colback=background!50,
        colframe=edge,
        width=\columnwidth,% total width
        boxrule = 0.3mm,
        top = 3pt, bottom=3pt, left=3pt, right=3pt
    ]
    \textbf{Answer to RQ4:} 
    \name can be adapted to multiple LLM families, but does not uniformly enhance every LLM. Its adaptation depends on the LLM's model family, as the representation spaces of some LLMs may not align well with that of the graph branch's embedding model.
\end{tcolorbox}

%%%% 消融实验
\subsection{RQ5: Contribution of Each Design in \name}

% The experimental results of the ablation study are shown in Table~\ref{tab:rq4}.
Table~\ref{tab:rq4} presents the experimental results of the ablation study.
\begin{table*}[htbp]
\centering
\caption{Experimental results of the ablation study}
\label{tab:rq4}
\resizebox{0.65\textwidth}{!}{
\begin{tabular}{cccccc}
\toprule
\textbf{Variant} & \textbf{Accuracy$\uparrow$} & \textbf{Precision$\uparrow$} & \textbf{Recall$\uparrow$} & \textbf{F1 Score$\uparrow$} & \textbf{FP Rate$\downarrow$} \\
\cmidrule(lr){1-6}\morecmidrules\cmidrule(lr){1-6}

\textit{w/o Graph branch} & 0.593 & 0.655 & 0.649 & 0.652 & 0.493 \\
\textit{w/o LLM branch} & 0.670 & 0.782 & 0.607 & 0.683 & \textbf{0.244} \\
\textit{w/o Adaptive fusion} & 0.649 & 0.665 & 0.810 & 0.730 & 0.580 \\
\textit{w/o Adapters} &  0.601 & 0.653 & 0.686 & 0.669 & 0.520 \\
\textit{w/o Gate network} & 0.676 & 0.679 & 0.851 & 0.755 & 0.573 \\
\textit{w/o Cross-attention} & 0.693 & 0.731 & 0.754 & 0.743 & 0.394 \\
\textit{w/o Two-stage training} & 0.702 & 0.706 & 0.845 & 0.769 & 0.501 \\

\midrule
\textbf{\name} & \textbf{0.854} & \textbf{0.823} & \textbf{0.957} & \textbf{0.885} & 0.293 \\

\bottomrule
\end{tabular}
}
\end{table*}

\textbf{\textit{Structure-Guided Joint Representation}}: 
The joint representation contributes the most to the effectiveness of \name, as removing the graph branch (w/o Graph branch) reduces accuracy by 30.56\% and removing the LLM branch (w/o LLM branch) reduces it by 21.55\%. Furthermore, the graph branch achieves a 68.26\% reduction in the false positive rate, demonstrating that structural information is essential for \name to identify security patches and avoid excessive false positives accurately. These results suggest that the structure-guided joint representation equips the LLM with critical structural cues while simultaneously leveraging its semantic understanding, thereby improving binary security patch detection.

% \lqy{
\textbf{\textit{Adaptive Fusion}}: 
However, a naive embedding-level joint representation, i.e., directly concatenating the GGCN graph embedding with the LLM’s last hidden state (w/o Adaptive fusion), substantially degrades performance, lowering accuracy by 24.00\% and F1 by 17.51\%, while increasing the false positive rate by 97.95\%.
We attribute this drop to a representation mismatch between the two modalities: the assembly CFG embedding and the pseudo-code embedding are not naturally aligned, and simple concatenation does not provide a mechanism to project them into a shared semantic space.
% }

\textbf{\textit{Modules in the Graph Branch}}:
Removing individual modules (w/o Adapters, w/o Gate, and w/o Cross-attention) in the graph branch also leads to notable accuracy drops of 29.63\%, 20.84\%, and 18.85\%, respectively. 
% \lqy{
In particular, the adapter yields a substantial performance gain for \name, corroborating our motivation in Section~\ref{section:method}: a simple linear projection is insufficient to map graph representations into the LLM’s latent space.
% }
Furthermore, the three modules are essential for enabling the LLM to accurately interpret non-security patches and reduce false positives, with reductions of 77.47\%, 95.56\%, and 34.47\% in the false positive rate, respectively.
% \lqy{
In particular, the gate network delivers a substantial reduction in the false positive rate for \name, also corroborating our motivation in Section~\ref{section:method}: it acts as a dynamic noise filter that effectively suppresses CFG topologies irrelevant to security patch patterns.
% }
These results highlight that equipping the LLM with fine-grained structural information not only substantially enhances SPD performance but also markedly decreases false positive predictions.

\begin{figure}[htbp]
  \centering

  \begin{minipage}[t]{\columnwidth}
    \centering
    \includegraphics[width=0.9\columnwidth]{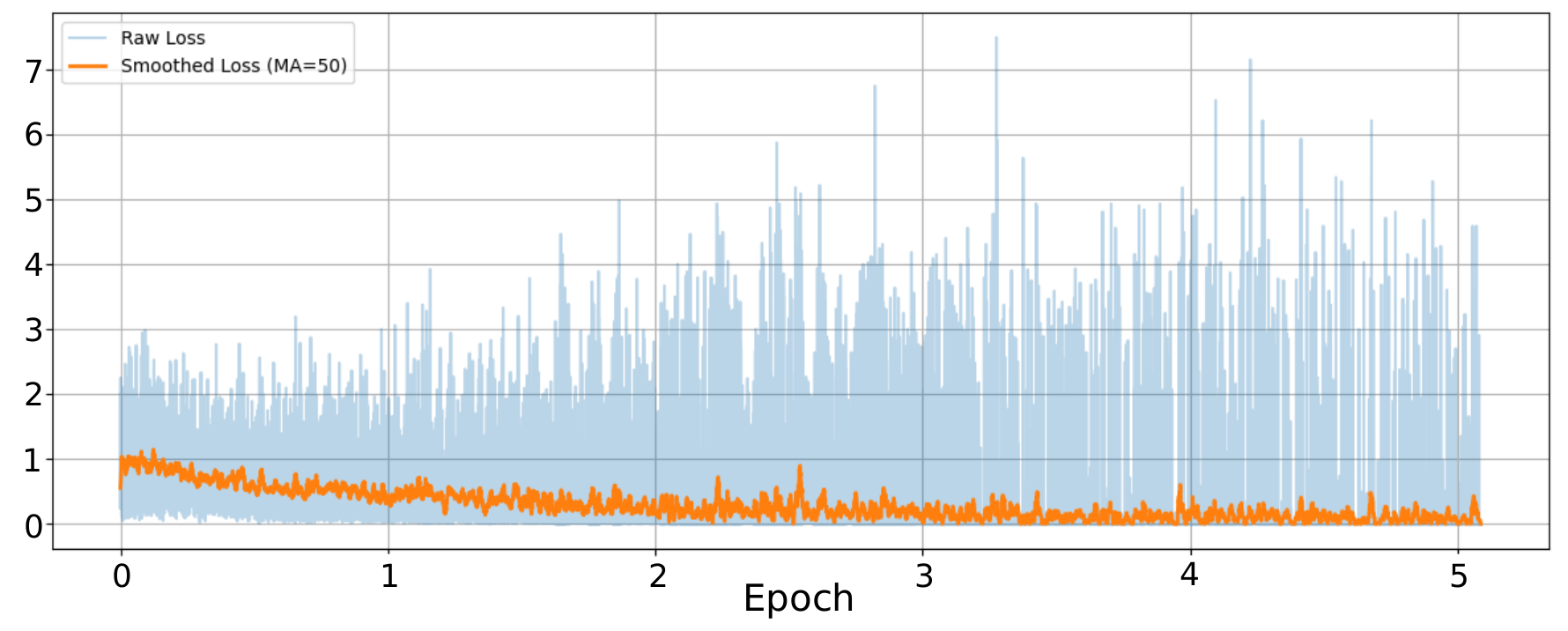}
  \end{minipage}

  \vspace{0.5em}

  \begin{minipage}[t]{\columnwidth}
    \centering
    \includegraphics[width=0.9\columnwidth]{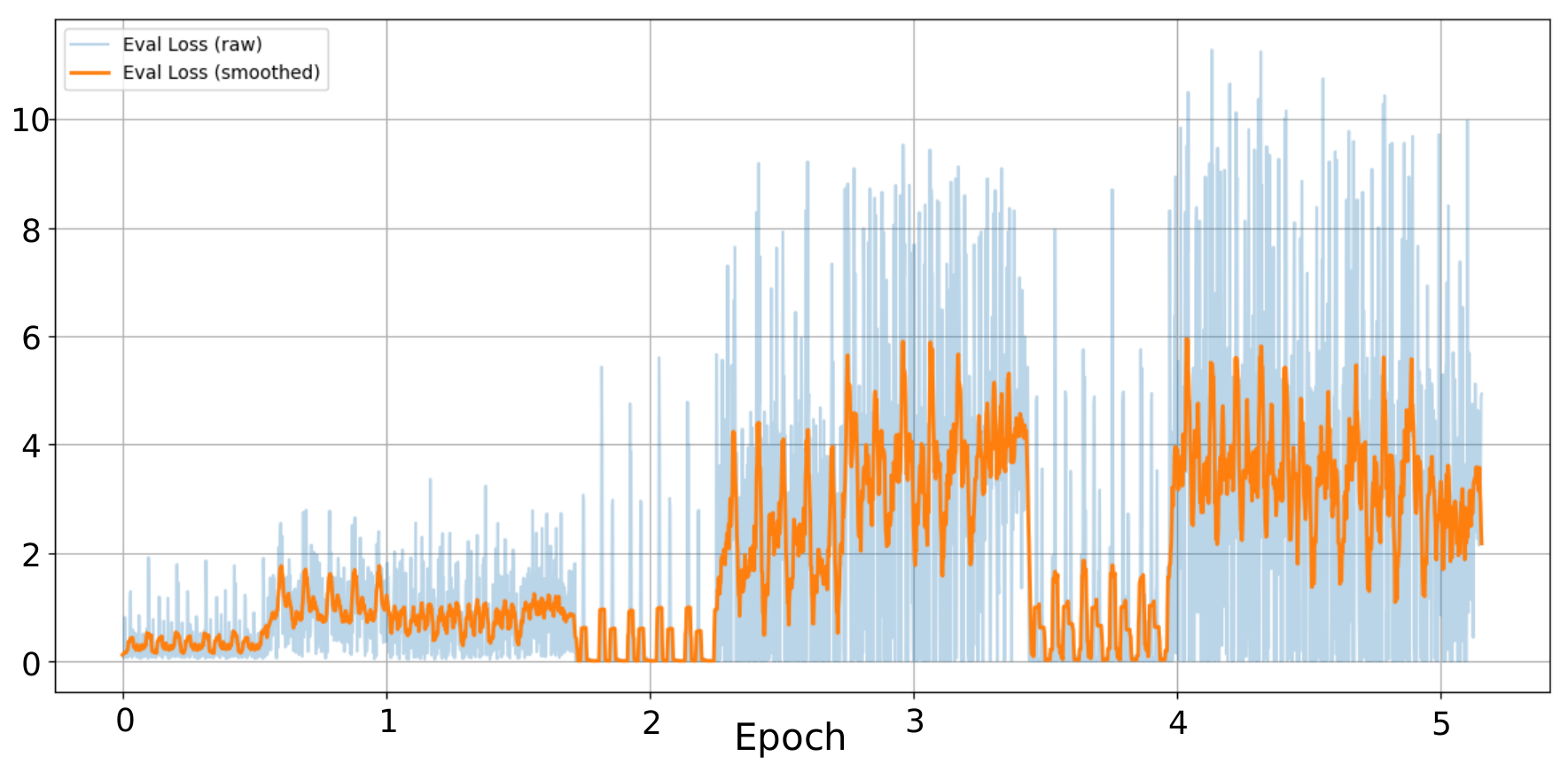}
  \end{minipage}

  \caption{Training and validation loss curves under joint training.}
  \label{fig:one_stage_losses}
\end{figure}

\textbf{\textit{Two-stage Training}}: 
Disabling the two-stage training strategy (w/o Two-stage training) decreases 14.22\% in precision and 13.11\% in F1 score. 
% \lqy{
To investigate why two-stage training is necessary, we track the optimization dynamics under joint training (w/o Two-stage training). The results reveal two failure modes: (1) although the training loss steadily decreased, the validation loss diverged sharply (spiking above 10 with an upward trend) as shown in Figure~\ref{fig:one_stage_losses}; and (2) the gradient cosine similarity between the two branches remained close to zero or became negative (e.g., $-9.9\times10^{-6}$ at epoch 3), suggesting that the branches often occur with misaligned update directions that hinder stable co-optimization.
% }

These result underscores that the graph branch and the LLM branch interfere with each other and cannot achieve balanced joint learning.

\begin{tcolorbox}[colback=background!50,
        colframe=edge,
        width=\columnwidth,% total width
        boxrule = 0.3mm,
        top = 3pt, bottom=3pt, left=3pt, right=3pt
    ]
    \textbf{Answer to RQ5:} 
    Each Design of \name, including the structure-guided joint representation, the modules in the graph branch, and the two-stage training strategy, effectively contributes to its effectiveness in accurately detecting security patches.
\end{tcolorbox}

% ===================================================================== %

% To investigate the effect of the graph branch on \name, we designed three experimental variants: (1) one variant removes the adapter module and instead employs simple linear layers to map the graph representation into the LLM’s semantic space(w/o Adapters), (2) another variant removes the gating mechanism, such that the graph representation, after being mapped by the adapter, is no longer selectively activated(w/o Gate), (3) the final variant eliminates the cross-attention module(w/o Cross-attention).

% To explore the impact of joint representation of assembly-level structure and pseudo-code semantics, we construct two experimental variants: (1) one solely utilizes the LLM branch(w/o Graph branch), (2) the other exclusively employs the graph branch(w/o LLM branch), which demonstrate the effectiveness of capturing structural information and semantic information, respectively.
\section{Discussion} \label{section:results}

\subsection{Qualitative Analysis}

\begin{figure}[t!]
\centering
\includegraphics[width=0.48\textwidth]{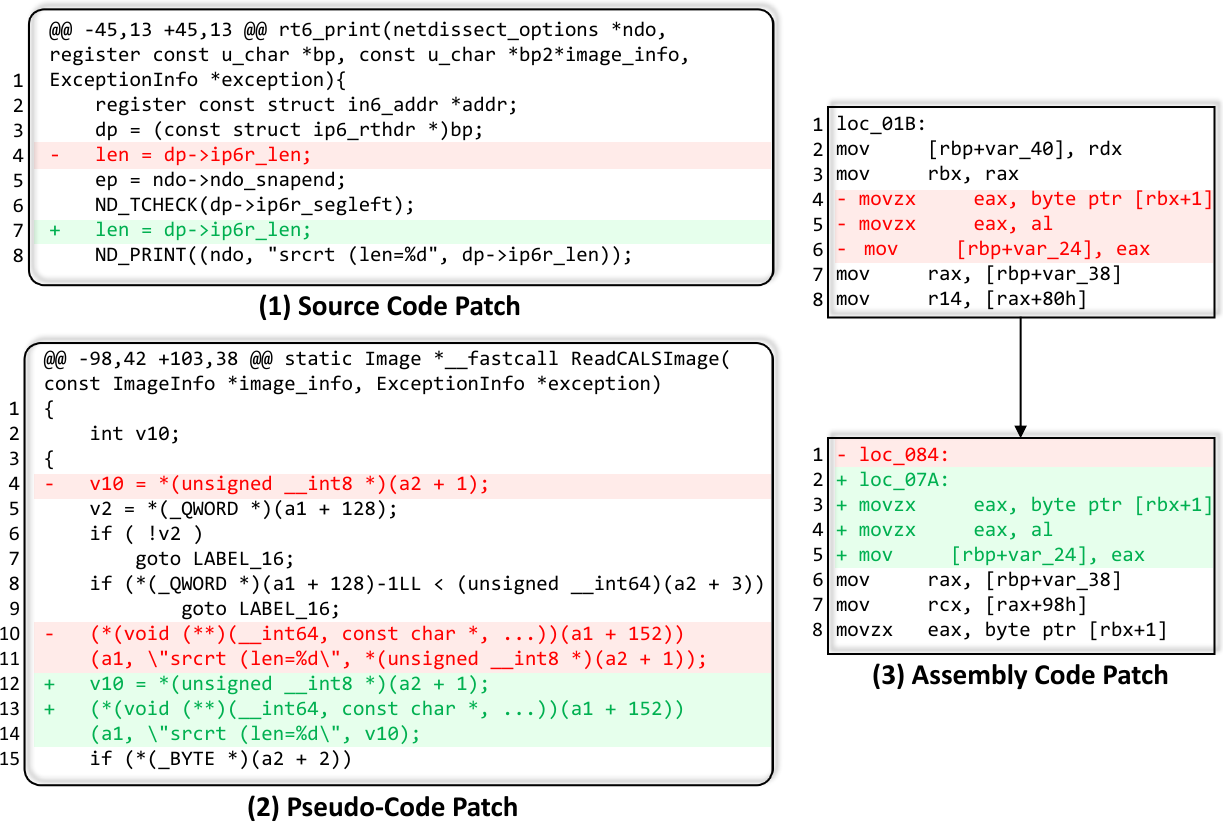}
\vspace{-0.5em}
\caption{Case study of the patch fixing CVE-2017-13725.}
\label{fig:case}
\vspace{-1em}
\end{figure}

% 1. 分析为什么我们的方法work
\textbf{\textit{Why does \name Work}}?
We identify the advantages of \name: GGCN captures fine-grained node-level and structural changes, and \textbf{adapters} effectively map those graph representations into the LLM’s hidden space. We analyze the ablation study results and identify a representative example that can only be detected by \name, while removing any module prevents the detection, thereby clearly illustrating why \name works.
This patch addresses a critical level buffer over-read vulnerability identified as CVE-2017-13725~\cite{cve201713725}, as shown in Figure~\ref{fig:case}. Insufficient validation of IPv6 routing header fields in \textit{ip6r\_len} causes out-of-bounds memory accesses, potentially leading to information disclosure or system crashes.
In Figure~\ref{fig:case}(1), by moving the buffer read operation \textit{len = dp \textrightarrow ip6r\_len} to occur after the existence check \textit{ND\_TCHECK(dp \textrightarrow ip6r\_segleft)}, this patch prevents a potential crash. 

In Figure~\ref{fig:case}(2), the buffer read operation is likewise shifted to a safe location in the pseudo-code patch. specifically, \textit{v10 = *(unsigned \_\_int8 *)(a2 + 1)} is moved to after \textit{if (!v2) goto LABEL\_16} and \textit{if (*(\_QWORD *)(a1 + 128) - 1LL < (unsigned \_\_int64)(a2 + 3)) goto LABEL\_16}.
However, the pseudo-code patch also changes a printing call as \textit{(*(void (**)(\_\_int64, const char *, ...))(a1 + 152))(a1, "srcrt (len=\%d", v10)} rather than the original inline form, which introduces a noise and significantly impairs the ability to identify this security patch using the pseudo-code patch alone.

In Figure~\ref{fig:case}(3), this operation appears as the movement of instructions at the assembly level: \textit{movzx eax, byte ptr [rbx+1]}, \textit{movzx eax, al}, and \textit{[rbp+var\_24], eax}, from the first basic block to the third basic block, reflecting a node-level change.
\name’s graph-branch captures this change and maps it into the LLM’s hidden representation space, effectively filtering out the semantic noise present in the pseudo-code and thereby enabling \name to identify this security patch.

% 2. 分析我们仍不能识别好的正负样本
\textbf{\textit{Error Analysis}}.
Despite \name’s superior performance in identifying security patches, it still fails in some cases.
We analyze 44 security patches in the buffer-overflow category that are frequently misidentified and find that most misidentifications result from the following causes:

\textbf{\textit{Missing Security-fix Patterns}}: The repair point depends on specific API semantics that may be absent from the training corpus. For example, the patch for CVE-2017-13051~\cite{cve201713051} adds a basic block to the CFG, but its instructions only operate on registers; such a structural change may not correspond to a directly mappable security-fix pattern in the training data.

\textbf{\textit{Insufficient CFG Changes}}: The fix produces almost no observable difference in the CFG, such as optimized instruction sequences remain identical, so the graph branch cannot detect the change. 
For example, the patch for CVE-2017-12897~\cite{cve201712897} shows almost no node or control-flow differences in the CFG, and the graph branch fails to capture node-level or structural changes.

\textit{\textbf{Cross-function Changes Obscure Semantics}}: The vulnerability-fixing logic is spread across multiple functions, meaning that one must jointly understand edits in several functions to determine whether a change constitutes a security patch. For example, in the patch for CVE-2023-1801~\cite{cve20231801}, the pseudo-code modifications occur not only in the current function but also in its callees. Consequently, the vulnerability-fixing logic is incomplete when viewed from this function alone.

\subsection{Limitations}
\textbf{\textit{Limited Generalizability to Some LLMs}}:
Although the results in Section~\ref{section:rq3} show that \name generalizes well to the Qwen, DeepSeek, and Llama families, it fails to adapt to some models in the Yi family. Notably, Yi-Coder-9B-Chat, the best-performing baseline, shows promising potential for adaptation.
In future work, we aim to investigate how to align the graph-branch embedding model with more LLMs, so as to further enhance \name’s generalizability to widely used LLMs across different parameter scales.

\textbf{\textit{Generalizability on Multiple Programming Languages}}: 
\name is specifically designed for C/C++, and accordingly, our experiments focus on C/C++. Nevertheless, \name can be extended to other languages such as Java, by extracting control-flow and data-flow information from Java bytecode to serve as input for the graph branch, while employing Java-like pseudo-code as input to the LLM branch.
In future work, we aim to investigate the applicability and effectiveness of \name across a broader range of programming languages.
% % lqy: 暂时先不添加这段。因为我没办法确认Java的伪代码是否和C的一样很难构建结构信息，如AST和CFG，而Java的bytecode则容易的多。如果没有这层约束，即Java-like pseudo-code很容易抽取结构信息，那我们的方案在Java上就不一定是理论上的最佳方案了，可以单从Java-like pseudo-code中同时获取结构和语义信息！！

\textbf{\textit{Restrictions of Decompilation}}:
\name leverages both representations of a binary patch to achieve strong performance.
However, every decompiler has a failure rate. This limitation is inherent in heuristic decompilers and the lossy nature of the compilation process (e.g., aggressive optimizations). 
A potential solution is to leverage LLMs specialized for decompilation~\cite{tan2024llm4decompile} to generate pseudo-code, which serves as input to \name.

\section{Threats to Validity}
The potential \textbf{internal threats} are as follows:

\textbf{\textit{Model Selection}}: 
We select Qwen3-8B as the backbone of \name, which may not be the optimal selection, as numerous open-source LLMs have recently demonstrated strong performance on code-related tasks. 
Nonetheless, the results in Section~\ref{section:rq3} demonstrate that Qwen3-8B achieves the best performance among many of the most popular LLM families, which mitigates this threat to some extent.

\textbf{\textit{Hyperparameter Settings}}:
The setting of hyperparameters can affect model performance. 
Therefore, we determine the hyperparameters for fine-tuning LLMs and training the graph branch based on prior studies~\cite{li2025empiricalstudycodelarge,wen2024repository}, and employ early stopping to prevent overfitting. 
These measures substantially mitigate the potential threat.

The potential \textbf{external threats} are as follows:

\textbf{\textit{Non-determinism in Large Language Model Outputs}}:
LLMs model the conditional probability distribution of input sequences, and their decoding relies on probabilistic sampling, which makes the outputs inherently non-deterministic.
However, they are not arbitrary, as they reflect the biases learned from extensive training data.
To minimize the threat introduced by non-determinism, we set \textit{top\_p}, \textit{top\_k}, and \textit{temperature} to 0 during inference with \name.

% 外部
% \textbf{\textit{Non-deterministic output of large language models}}.
% The settings for \textit{top\_k} and \textit{temperature} induce non-deterministic outputs in LLMs. Nevertheless, these outputs are not entirely random, as outputs conform to the biases that LLMs learn from extensive training data. We employ the default inference method provided by LLaMAFactory to ensure that the LLM's output remains as stable as possible.

\section{Related Work} \label{section:related}

% lqy: 新的写作思路，有时间了替换。

% 1. decompilition：介绍经典的反编译工作/工具，以及他们的优点与不足，以服务于为什么会有assembly code和pseudo-code这两种抽象表示。
\subsection{Decompiler and Decompilation}

Decompilers play a crucial role in binary analysis by lifting machine code into higher-level representations, such as assembly or pseudo-code. They provide critical high-level representations for binary security patch detection.

\textbf{\textit{IDA Pro}}~\cite{idapro} is a widely used commercial decompiler supporting multiple processor architectures, including x86 and ARM.
Its core lies in instruction set mapping and control-flow-driven parsing. Through a built-in instruction set database, it maps raw byte streams to assembly instructions and employs two complementary algorithms, linear scanning and recursive descent, to distinguish code from data.
Its decompilation primarily relies on the Hex-Rays plugin, which converts assembly instructions into a custom microcode intermediate representation and generates C-like pseudo-code by combining control flow analysis and data flow tracking.

\textbf{\textit{Ghidra}}~\cite{ghidra} is an open-source decompilation framework developed by the National Security Agency (NSA).
Its disassembly process maps binaries to assembly code by matching instruction rules sequentially along byte streams, based on predefined opcode meanings and addressing modes.
Its decompilation module first converts assembly instructions into a custom P-code intermediate representation through a rule-driven manner, then generates pseudo-code through heuristic control flow recovery and type inference.

% \textbf{\textit{Binary Ninja}}~\cite{binaryninja} is an interactive binary analysis platform with extensive community extensions. 
% The core of its disassembly lies in mapping machine code to assembly instructions by rapidly matching entries in predefined tables, using a parsing strategy that employs recursive descent by default but automatically switches to linear scanning when encountering obfuscated code (e.g., flattened control flow).
% Its decompilation process first converts assembly instructions into a three-level intermediate language (i.e., low-, medium-, and high-level IL), then incorporates heuristic rules to generate structured pseudo-code, and ultimately employs mutation operators to reconstruct obfuscated code.

% 2. LLM 4 Binary-related work 服务于我们使用LLM4BinarySPD
\subsection{Binary Task-specific LLM}

In recent years, research advances have demonstrated the remarkable capabilities of LLMs in several binary-code-related tasks, including decompilation, compilation optimization, and binary code similarity detection.

\textbf{\textit{LLM-Compiler}}~\cite{cummins2025llm} family is a foundation language model series designed by Meta~\cite{meta} based on CodeLlama~\cite{codellama}, specifically for binary-related tasks. 
This family includes \textit{LLM-Compiler-7B} and \textit{LLM-Compiler-7B-ftd}, which are state-of-the-art binary code models with enhanced capabilities in assembly code processing, optimization, and compiler reasoning.
The base LLM-Compiler-7B is pre-trained on over 500 billion tokens spanning x86\_64, ARM, and CUDA assembly as well as LLVM-IR, enabling it to predict the effects of compiler optimizations and emulate compiler behavior. The ftd variant (i.e., fine-tuned for disassembly) further specializes in two critical tasks: optimizing LLVM assembly for code size and disassembling assembly to LLVM-IR.

\textbf{\textit{LLM4Decompile}}~\cite{tan2024llm4decompile} family is another series of foundation language models designed by \citet{tan2024llm4decompile}, built on DeepSeek-Coder~\cite{guo2024deepseek} and Yi-Coder~\cite{yi-coder} for decompilation tasks.
\textit{LLM4Decompile-9B-v2} is a state-of-the-art binary code model that targets decompilation from assembly to C source. As an open-source model built on the Yi-Coder-9B architecture, it specializes in converting Linux x86\_64 binaries into human-readable and executable C code. Its post-training focuses on two core decompilation tasks: (1) End-to-end translation for direct decompilation of assembly code; (2) Refinement of pseudo-code lifted from traditional decompilers (e.g., Ghidra).
Notably, it achieves a re-executability rate of 0.6494 on the Decompile benchmark, significantly outperforming GPT-4o and Ghidra in terms of code executability and readability.

% \textbf{\textit{Nova}}~\cite{jiang2025nova} family is a series of large language models designed for assembly code optimization and binary code decompilation.
% \textit{Nova-6.7B} is pre-trained with a language modeling objective, initialized from DeepSeek-Coder. Its training data includes disassembly code from AnghaBench and C/C++ programs compiled from The-Stack. 
% Notably, it outperforms existing techniques in binary code decompilation by up to 14.84\% in Pass@1, and surpasses the latest binary code similarity detection techniques by 6.17\% in recall. 

% 3. binary-code-related tasks
\subsection{Binary Analysis}

Binary analysis provides insight into compiled programs when source code is unavailable and it is essential for many security tasks, such as binary security patch identification, vulnerability detection, and taint analysis. Recent advances in deep learning have enabled substantial progress in these binary-based tasks.

% \textbf{\textit{BinPool}}~\cite{arasteh2025binpool} is a publicly available dataset of real-world binary vulnerabilities constructed by correlating NVD records with Debian security data and automatically building vulnerable and patched binary pairs. The collection covers 603 CVEs across 89 CWE categories, 162 Debian packages, and 6,144 binaries compiled at four optimization levels, and it includes fine-grained metadata linking patches to files, functions, and source/binary locations. Its automated pipeline and matched binaries make it useful for evaluating both DL-based detectors and traditional program-analysis tools.

\textbf{\textit{BinMetric}}~\cite{shang2025binmetric} is a large-scale and comprehensive binary code analysis benchmark of 1,000 problems from 20 open-source projects. It was introduced by \citet{shang2025binmetric} for empirically evaluating large language models on 6 practical binary-code-related tasks, which reflect actual reverse engineering scenarios. The strengths and limitations of various state-of-the-art LLMs were investigates in this study via an automated multi-evaluator pipeline. The results show that while LLMs show strong potential, but still struggle with precise binary lifting and assembly synthesis.

\textbf{\textit{LATTE}}~\cite{liu2023harnessing} is a static binary-taint analysis framework powered by large language models. It leverages LLMs to identify sinks and to locate potentially vulnerable destinations (i.e., call sites of security-sensitive functions) that warrant inspection. For each identified destination, LATTE performs backward intra- and inter-procedural data-dependency slicing to construct function-call chains that trace how tainted data can reach the vulnerable call sites. According to the authors, LATTE’s vulnerability-checking accuracy and F1 score surpass current state-of-the-art methods across all evaluated vulnerability types, and it notably achieves 100\% coverage for identifying sinks and sources in the tested scenario.

\section{Conclusion} \label{section:conclusion}

In this paper, we propose \name, a binary SPD framework that comprises: (1) a structure-guided neural network, which integrates an LLM branch and a graph branch to construct a joint representation of assembly code and pseudo-code, and (2) a two-stage training strategy, which mitigates the optimization imbalance caused by the significant disparity in parameter scale between the two branches and accelerates the training process.
To evaluate \name under realistic closed-source conditions, where patches from the same project or even the same application domain as the target binary are typically inaccessible, we construct a benchmark that is disjoint from the previous datasets in both project and domain. 
This benchmark comprises \textbf{1,720} entries drawn from five projects spanning five distinct domains.
\name achieves superior performance on this cross-project and cross-domain benchmark, outperforming the state-of-the-art baseline \textbf{12.66\%} in accuracy, and underscoring its effectiveness in real-world binary SPD.

% \name encodes the structural information from assembly-code patches by the graph branch to integrate it into the LLM branch, and leverages the LLM's semantic understanding to capture the semantics of pseudo-code patches, jointly representing both structural and semantic information of binary patches. Furthermore, we leverage the two-stage training algorithm to optimize \name, mitigating the issue of the significant parameter disparity between the LLM and graph branches.

% \input{sections/10_Others}

\bibliographystyle{IEEEtranN}
\bibliography{citation}

@String{Computing = "Computing" }

@String{Computer = "{IEEE} Computer" }

@String{Springer = "Springer-Verlag" }

@article{li2025empiricalstudycodelarge,
      title={Empirical Study of Code Large Language Models for Binary Security Patch Detection}, 
      author={Qingyuan Li and Binchang Li and Cuiyun Gao and Shuzheng Gao and Zongjie Li},
      journal={arXiv preprint arXiv:2509.06052},
      year={2025}
}

@inproceedings{wang2021patchrnn,
  title={Patchrnn: A deep learning-based system for security patch identification},
  author={Wang, Xinda and Wang, Shu and Feng, Pengbin and Sun, Kun and Jajodia, Sushil and Benchaaboun, Sanae and Geck, Frank},
  booktitle={MILCOM 2021-2021 IEEE Military Communications Conference (MILCOM)},
  pages={595--600},
  year={2021},
  organization={IEEE}
}

@inproceedings{he2024bingo,
  title={Bingo: Identifying security patches in binary code with graph representation learning},
  author={He, Xu and Wang, Shu and Feng, Pengbin and Wang, Xinda and Sun, Shiyu and Li, Qi and Sun, Kun},
  booktitle={Proceedings of the 19th ACM Asia Conference on Computer and Communications Security},
  pages={1186--1199},
  year={2024}
}

@article{tang2023just,
  title={Just-in-Time Security Patch Detection--LLM At the Rescue for Data Augmentation},
  author={Tang, Xunzhu and Chen, Zhenghan and Kim, Kisub and Tian, Haoye and Ezzini, Saad and Klein, Jacques},
  journal={arXiv preprint arXiv:2312.01241},
  year={2023}
}

@article{wen2024repository,
  title={Repository-Level Graph Representation Learning for Enhanced Security Patch Detection},
  author={Wen, Xin-Cheng and Lin, Zirui and Gao, Cuiyun and Zhang, Hongyu and Wang, Yong and Liao, Qing},
  journal={arXiv preprint arXiv:2412.08068},
  year={2024}
}

@article{hu2022lora,
  title={Lora: Low-rank adaptation of large language models.},
  author={Hu, Edward J and Shen, Yelong and Wallis, Phillip and Allen-Zhu, Zeyuan and Li, Yuanzhi and Wang, Shean and Wang, Lu and Chen, Weizhu and others},
  journal={ICLR},
  volume={1},
  number={2},
  pages={3},
  year={2022}
}

@inproceedings{wang2024reposvul,
  title={Reposvul: A repository-level high-quality vulnerability dataset},
  author={Wang, Xinchen and Hu, Ruida and Gao, Cuiyun and Wen, Xin-Cheng and Chen, Yujia and Liao, Qing},
  booktitle={Proceedings of the 2024 IEEE/ACM 46th International Conference on Software Engineering (ICSE): Companion Proceedings},
  pages={472--483},
  year={2024}
}

@inproceedings{wang2021patchdb,
  title={Patchdb: A large-scale security patch dataset},
  author={Wang, Xinda and Wang, Shu and Feng, Pengbin and Sun, Kun and Jajodia, Sushil},
  booktitle={2021 51st Annual IEEE/IFIP International Conference on Dependable Systems and Networks (DSN)},
  pages={149--160},
  year={2021},
  organization={IEEE}
}

@article{qwen2,
  title={Qwen2 technical report},
  author={Team, Qwen},
  journal={arXiv preprint arXiv:2407.10671},
  year={2024}
}

@article{qwen2.5coder,
  author={Hui, Binyuan and Yang, Jian and Cui, Zeyu and Yang, Jiaxi and Liu, Dayiheng and Zhang, Lei and Liu, Tianyu and Zhang, Jiajun and Yu, Bowen and Lu, Keming and others},
  title={Qwen2.5-Coder technical report},
  journal={arXiv preprint arXiv:2409.12186},
  year={2024}
}

@article{deepseekllm,
  title={Deepseek llm: Scaling open-source language models with longtermism},
  author={Bi, Xiao and Chen, Deli and Chen, Guanting and Chen, Shanhuang and Dai, Damai and Deng, Chengqi and Ding, Honghui and Dong, Kai and Du, Qiushi and Fu, Zhe and others},
  journal={arXiv preprint arXiv:2401.02954},
  year={2024}
}

@article{guo2024deepseek,
  title={DeepSeek-Coder: When the Large Language Model Meets Programming--The Rise of Code Intelligence},
  author={Guo, Daya and Zhu, Qihao and Yang, Dejian and Xie, Zhenda and Dong, Kai and Zhang, Wentao and Chen, Guanting and Bi, Xiao and Wu, Yu and Li, YK and others},
  journal={arXiv preprint arXiv:2401.14196},
  year={2024}
}

@article{deepseekcoderv2,
  title={Deepseek-coder-v2: Breaking the barrier of closed-source models in code intelligence},
  author={Zhu, Qihao and Guo, Daya and Shao, Zhihong and Yang, Dejian and Wang, Peiyi and Xu, Runxin and Wu, Y and Li, Yukun and Gao, Huazuo and Ma, Shirong and others},
  journal={arXiv preprint arXiv:2406.11931},
  year={2024}
}

@article{codellama,
  title={Code llama: Open foundation models for code},
  author={Roziere, Baptiste and Gehring, Jonas and Gloeckle, Fabian and Sootla, Sten and Gat, Itai and Tan, Xiaoqing Ellen and Adi, Yossi and Liu, Jingyu and Sauvestre, Romain and Remez, Tal and others},
  journal={arXiv preprint arXiv:2308.12950},
  year={2023}
}

@article{yi-coder,
  title={Yi: Open foundation models by 01. ai},
  author={Young, Alex and Chen, Bei and Li, Chao and Huang, Chengen and Zhang, Ge and Zhang, Guanwei and Wang, Guoyin and Li, Heng and Zhu, Jiangcheng and Chen, Jianqun and others},
  journal={arXiv preprint arXiv:2403.04652},
  year={2024}
}

@inproceedings{cummins2025llm,
  title={LLM Compiler: Foundation Language Models for Compiler Optimization},
  author={Cummins, Chris and Seeker, Volker and Grubisic, Dejan and Roziere, Baptiste and Gehring, Jonas and Synnaeve, Gabriel and Leather, Hugh},
  booktitle={Proceedings of the 34th ACM SIGPLAN International Conference on Compiler Construction},
  pages={141--153},
  year={2025}
}

@inproceedings{tan2024llm4decompile,
  title={LLM4Decompile: Decompiling Binary Code with Large Language Models},
  author={Tan, Hanzhuo and Luo, Qi and Li, Jing and Zhang, Yuqun},
  booktitle={Proceedings of the 2024 Conference on Empirical Methods in Natural Language Processing (EMNLP)},
  pages={3473--3487},
  year={2024}
}

@inproceedings{kipf2016gcn,
  author    = {Thomas N. Kipf and
               Max Welling},
  title     = {Semi-Supervised Classification with Graph Convolutional Networks},
  booktitle = {5th International Conference on Learning Representations, {ICLR} 2017,
               Toulon, France, April 24-26, 2017, Conference Track Proceedings},
  year      = {2017}
}

@inproceedings{xu2017spain,
  title={Spain: security patch analysis for binaries towards understanding the pain and pills},
  author={Xu, Zhengzi and Chen, Bihuan and Chandramohan, Mahinthan and Liu, Yang and Song, Fu},
  booktitle={2017 IEEE/ACM 39th International Conference on Software Engineering (ICSE)},
  pages={462--472},
  year={2017},
  organization={IEEE}
}

@article{achiam2023gpt,
  title={Gpt-4 technical report},
  author={Achiam, Josh and Adler, Steven and Agarwal, Sandhini and Ahmad, Lama and Akkaya, Ilge and Aleman, Florencia Leoni and Almeida, Diogo and Altenschmidt, Janko and Altman, Sam and Anadkat, Shyamal and others},
  journal={arXiv preprint arXiv:2303.08774},
  year={2023}
}

@inproceedings{lang2021pmatch,
  title={PMatch: Semantic-based patch detection for binary programs},
  author={Lang, Zhe and Yang, Shouguo and Cheng, Yiran and Zhang, Xiaoling and Shi, Zhiqiang and Sun, Limin},
  booktitle={2021 IEEE International Performance, Computing, and Communications Conference (IPCCC)},
  pages={1--10},
  year={2021},
  organization={IEEE}
}

@inproceedings{peng20191dvul,
  title={1dvul: Discovering 1-day vulnerabilities through binary patches},
  author={Peng, Jiaqi and Li, Feng and Liu, Bingchang and Xu, Lili and Liu, Binghong and Chen, Kai and Huo, Wei},
  booktitle={2019 49th Annual IEEE/IFIP International Conference on Dependable Systems and Networks (DSN)},
  pages={605--616},
  year={2019},
  organization={IEEE}
}

@article{economides2006two,
  title={Two-sided competition of proprietary vs. open source technology platforms and the implications for the software industry},
  author={Economides, Nicholas and Katsamakas, Evangelos},
  journal={Management science},
  volume={52},
  number={7},
  pages={1057--1071},
  year={2006},
  publisher={Informs}
}

@article{lee2005open,
  title={Open source vs. proprietary software: competition and compatibility},
  author={Lee, Sang-Yong Tom and Meng, Zhaoli},
  journal={Proprietary Software: Competition and Compatibility},
  year={2005}
}

@inproceedings{li2017large,
  title={A large-scale empirical study of security patches},
  author={Li, Frank and Paxson, Vern},
  booktitle={Proceedings of the 2017 ACM SIGSAC Conference on Computer and Communications Security (CCS)},
  pages={2201--2215},
  year={2017}
}

@article{NIPS2017_3f5ee243,
 title = {Attention is All You Need},
 author = {Vaswani, Ashish and Shazeer, Noam and Parmar, Niki and Uszkoreit, Jakob and Jones, Llion and Gomez, Aidan N and Kaiser, \L ukasz and Polosukhin, Illia},
 journal = {Advances in Neural Information Processing Systems (NeurIPS)},
 volume = {30},
 pages={5998--6008},
 year = {2017}
}

@article{yang2025qwen3,
  title={Qwen3 technical report},
  author={Yang, An and Li, Anfeng and Yang, Baosong and Zhang, Beichen and Hui, Binyuan and Zheng, Bo and Yu, Bowen and Gao, Chang and Huang, Chengen and Lv, Chenxu and others},
  journal={arXiv preprint arXiv:2505.09388},
  year={2025}
}

@article{cifuentes1995decompilation,
  title={Decompilation of binary programs},
  author={Cifuentes, Cristina and Gough, K John},
  journal={Software: Practice and Experience (SPE)},
  volume={25},
  number={7},
  pages={811--829},
  year={1995},
  publisher={Wiley Online Library}
}

@inproceedings{cao2024evaluating,
  title={Evaluating the effectiveness of decompilers},
  author={Cao, Ying and Zhang, Runze and Liang, Ruigang and Chen, Kai},
  booktitle={Proceedings of the 33rd ACM SIGSOFT International Symposium on Software Testing and Analysis (ISSTA)},
  pages={491--502},
  year={2024}
}

@article{cooper2002building,
  title={Building a control-flow graph from scheduled assembly code},
  author={Cooper, Keith D and Harvey, Timothy J and Waterman, Todd},
  journal={Dept. of Computer Science, Rice University},
  year={2002}
}

@inproceedings{brumley2013native,
  title={Native x86 decompilation using Semantics-Preserving structural analysis and iterative Control-Flow structuring},
  author={Brumley, David and Lee, JongHyup and Schwartz, Edward J and Woo, Maverick},
  booktitle={22nd USENIX Security Symposium (USENIX Security 13)},
  pages={353--368},
  year={2013}
}

@inproceedings{burk2022decomperson,
  title={Decomperson: How humans decompile and what we can learn from it},
  author={Burk, Kevin and Pagani, Fabio and Kruegel, Christopher and Vigna, Giovanni},
  booktitle={31st USENIX Security Symposium (USENIX Security 22)},
  pages={2765--2782},
  year={2022}
}

@inproceedings{guographcodebert,
  title={GraphCodeBERT: Pre-training Code Representations with Data Flow},
  author={Guo, Daya and Ren, Shuo and Lu, Shuai and Feng, Zhangyin and Tang, Duyu and LIU, Shujie and Zhou, Long and Duan, Nan and Svyatkovskiy, Alexey and Fu, Shengyu and others},
  booktitle={International Conference on Learning Representations (ICLR)},
  year={2021}
}

@inproceedings{guo2022unixcoder,
  title={UniXcoder: Unified Cross-Modal Pre-training for Code Representation},
  author={Guo, Daya and Lu, Shuai and Duan, Nan and Wang, Yanlin and Zhou, Ming and Yin, Jian},
  booktitle={Proceedings of the 60th Annual Meeting of the Association for Computational Linguistics (Volume 1: Long Papers)},
  pages={7212--7225},
  year={2022}
}

@inproceedings{niu2024fair,
  title={Fair: Flow type-aware pre-training of compiler intermediate representations},
  author={Niu, Changan and Li, Chuanyi and Ng, Vincent and Lo, David and Luo, Bin},
  booktitle={Proceedings of the 46th IEEE/ACM International Conference on Software Engineering (ICSE)},
  pages={1--12},
  year={2024}
}

@article{touvron2023llama,
  title={Llama: Open and efficient foundation language models},
  author={Touvron, Hugo and Lavril, Thibaut and Izacard, Gautier and Martinet, Xavier and Lachaux, Marie-Anne and Lacroix, Timoth{\'e}e and Rozi{\`e}re, Baptiste and Goyal, Naman and Hambro, Eric and Azhar, Faisal and others},
  journal={arXiv preprint arXiv:2302.13971},
  year={2023}
}

@article{shengyu2023instruction,
  title={Instruction tuning for large language models: A survey},
  author={Shengyu, Zhang and Linfeng, Dong and Xiaoya, Li and Sen, Zhang and Xiaofei, Sun and Shuhe, Wang and Jiwei, Li and Hu, Runyi and Tianwei, Zhang and Wu, Fei and others},
  journal={arXiv preprint arXiv:2308.10792},
  year={2023}
}

@article{wu2021representing,
  title={Representing long-range context for graph neural networks with global attention},
  author={Wu, Zhanghao and Jain, Paras and Wright, Matthew and Mirhoseini, Azalia and Gonzalez, Joseph E and Stoica, Ion},
  journal={Advances in neural information processing systems (NeurIPS)},
  volume={34},
  pages={13266--13279},
  year={2021}
}

@article{bresson2017residual,
  title={Residual gated graph convnets},
  author={Bresson, Xavier and Laurent, Thomas},
  journal={arXiv preprint arXiv:1711.07553},
  year={2017}
}

@incollection{kruse2022multi,
  title={Multi-layer perceptrons},
  author={Kruse, Rudolf and Mostaghim, Sanaz and Borgelt, Christian and Braune, Christian and Steinbrecher, Matthias},
  booktitle={Computational intelligence: a methodological introduction},
  pages={53--124},
  year={2022},
  publisher={Springer}
}

@article{shang2025binmetric,
  title={BinMetric: A Comprehensive Binary Analysis Benchmark for Large Language Models},
  author={Shang, Xiuwei and Chen, Guoqiang and Cheng, Shaoyin and Wu, Benlong and Hu, Li and Li, Gangyang and Zhang, Weiming and Yu, Nenghai},
  journal={arXiv preprint arXiv:2505.07360},
  year={2025}
}

@inproceedings{kruegel2004static,
  title={Static disassembly of obfuscated binaries},
  author={Kruegel, Christopher and Robertson, William and Valeur, Fredrik and Vigna, Giovanni},
  booktitle={USENIX security Symposium},
  volume={13},
  pages={18--18},
  year={2004}
}

@inproceedings{avgerinos2011tie,
  title={TIE: Principled reverse engineering of types in binary programs},
  author={Avgerinos, Thanassis},
  booktitle={Proceedings of the 18th Annual Network and Distributed System Security Symposium (NDSS)},
  year={2011}
}

@article{banerjee2021variable,
  title={Variable name recovery in decompiled binary code using constrained masked language modeling},
  author={Banerjee, Pratyay and Pal, Kuntal Kumar and Wang, Fish and Baral, Chitta},
  journal={arXiv preprint arXiv:2103.12801},
  year={2021}
}

@inproceedings{chen2023investigating,
  title={Investigating neural-based function name reassignment from the perspective of binary code representation},
  author={Chen, Guoqiang and Gao, Han and Zhang, Jie and He, Yanru and Cheng, Shaoyin and Zhang, Weiming},
  booktitle={2023 20th Annual International Conference on Privacy, Security and Trust (PST)},
  pages={1--11},
  year={2023},
  organization={IEEE}
}

@article{telang2007empirical,
  title={An empirical analysis of the impact of software vulnerability announcements on firm stock price},
  author={Telang, Rahul and Wattal, Sunil},
  journal={IEEE Transactions on Software Engineering (TSE)},
  volume={33},
  number={8},
  pages={544--557},
  year={2007},
  publisher={IEEE}
}

@inproceedings{linn2003obfuscation,
  title={Obfuscation of executable code to improve resistance to static disassembly},
  author={Linn, Cullen and Debray, Saumya},
  booktitle={Proceedings of the 10th ACM conference on Computer and communications security (CCS)},
  pages={290--299},
  year={2003}
}

@article{klabunde2023towards,
  title={Towards measuring representational similarity of large language models},
  author={Klabunde, Max and Amor, Mehdi Ben and Granitzer, Michael and Lemmerich, Florian},
  journal={arXiv preprint arXiv:2312.02730},
  year={2023}
}

@inproceedings{dai2024representational,
  title={Representational Analysis of Binding in Language Models},
  author={Dai, Qin and Heinzerling, Benjamin and Inui, Kentaro},
  booktitle={Proceedings of the 2024 Conference on Empirical Methods in Natural Language Processing (EMNLP)},
  pages={17468--17493},
  year={2024}
}

@article{liu2023harnessing,
  title={Harnessing the power of llm to support binary taint analysis},
  author={Liu, Puzhuo and Sun, Chengnian and Zheng, Yaowen and Feng, Xuan and Qin, Chuan and Wang, Yuncheng and Li, Zhi and Sun, Limin},
  journal={arXiv preprint arXiv:2310.08275},
  year={2023}
}

@article{aghajanyan2020intrinsic,
  title={Intrinsic dimensionality explains the effectiveness of language model fine-tuning},
  author={Aghajanyan, Armen and Zettlemoyer, Luke and Gupta, Sonal},
  journal={arXiv preprint arXiv:2012.13255},
  year={2020}
}

@inproceedings{glavavs2020non,
  title={Non-linear instance-based cross-lingual mapping for non-isomorphic embedding spaces},
  author={Glava{\v{s}}, Goran and Vuli{\'c}, Ivan},
  booktitle={Proceedings of the 58th annual meeting of the association for computational linguistics},
  pages={7548--7555},
  year={2020}
}

@article{rumelhart1986learning,
  title={Learning representations by back-propagating errors},
  author={Rumelhart, David E and Hinton, Geoffrey E and Williams, Ronald J},
  journal={nature},
  volume={323},
  number={6088},
  pages={533--536},
  year={1986},
  publisher={Nature Publishing Group UK London}
}

@inproceedings{arakawa2025towards,
  title={Towards the Identification of Vulnerability-Fixing Code Lines in OSS Security Patches Using Lexical Code Segmentation and LLMs},
  author={Arakawa, Reika Nishimura and Kanemoto, Yo and Akiyama, Mitsuaki},
  booktitle={IFIP Annual Conference on Data and Applications Security and Privacy},
  pages={73--95},
  year={2025},
  organization={Springer}
}

@inproceedings{sun2025dispatch,
  title={DISPATCH: Unraveling Security Patches from Entangled Code Changes},
  author={Sun, Shiyu and Xing, Yunlong and Wang, Xinda and Wang, Shu and Li, Qi and Sun, Kun},
  booktitle={34th USENIX Security Symposium (USENIX Security 25)},
  pages={4521--4540},
  year={2025}
}

@inproceedings{luo2024strengthening,
  title={Strengthening supply chain security with fine-grained safe patch identification},
  author={Luo, Changhua and Meng, Wei and Wang, Shuai},
  booktitle={Proceedings of the IEEE/ACM 46th International Conference on Software Engineering (ICSE)},
  pages={1--12},
  year={2024}
}

@inproceedings{liu2020far,
  title={How far we have come: Testing decompilation correctness of C decompilers},
  author={Liu, Zhibo and Wang, Shuai},
  booktitle={Proceedings of the 29th ACM SIGSOFT International Symposium on Software Testing and Analysis (ISSTA)},
  pages={475--487},
  year={2020}
}

@inproceedings{wang2024improving,
  title={Improving ML-based Binary Function Similarity Detection by Assessing and Deprioritizing Control Flow Graph Features},
  author={Wang, Jialai and Zhang, Chao and Chen, Longfei and Rong, Yi and Wu, Yuxiao and Wang, Hao and Tan, Wende and Li, Qi and Li, Zongpeng},
  booktitle={33rd USENIX Security Symposium (USENIX Security 24)},
  pages={4265--4282},
  year={2024}
}

@inproceedings{xu2017neural,
  title={Neural network-based graph embedding for cross-platform binary code similarity detection},
  author={Xu, Xiaojun and Liu, Chang and Feng, Qian and Yin, Heng and Song, Le and Song, Dawn},
  booktitle={Proceedings of the 2017 ACM SIGSAC conference on computer and communications security},
  pages={363--376},
  year={2017}
}

@inproceedings{stephens2016driller,
  title={Driller: Augmenting fuzzing through selective symbolic execution.},
  author={Stephens, Nick and Grosen, John and Salls, Christopher and Dutcher, Andrew and Wang, Ruoyu and Corbetta, Jacopo and Shoshitaishvili, Yan and Kruegel, Christopher and Vigna, Giovanni},
  booktitle={NDSS},
  volume={16},
  number={2016},
  pages={1--16},
  year={2016}
}

@inproceedings{andriesse2016depth,
  title={An In-Depth analysis of disassembly on Full-Scale x86/x64 binaries},
  author={Andriesse, Dennis and Chen, Xi and Van Der Veen, Victor and Slowinska, Asia and Bos, Herbert},
  booktitle={25th USENIX security symposium (USENIX security 16)},
  pages={583--600},
  year={2016}
}

@inproceedings{kornblith2019similarity,
  title={Similarity of neural network representations revisited},
  author={Kornblith, Simon and Norouzi, Mohammad and Lee, Honglak and Hinton, Geoffrey},
  booktitle={International conference on machine learning (ICML)},
  pages={3519--3529},
  year={2019},
  organization={PMlR}
}

@misc{EICReport,
  author = {McAfee \& CSIS},
  title = {Economic Impact of Cybercrime—No Slowing Down},
  year = {2018},
  howpublished  ={[Online]. Available: https://csis-website-prod.s3.amazonaws.com/s3fs-public/publication/economic-impact-cybercrime.pdf}
}

@misc{OssraReport,
  author = {BlackDuck},
  title = {2025 Open Source Security and Risk Analysis Report},
  year = {2025},
  howpublished  ={[Online]. Available: https://www.blackduck.com/content/dam/black-duck/en-us/reports/rep-ossra.pdf}
}

@misc{Linux,
  author = {Linux},
  year = {2025},
  howpublished  ={[Online]. Available: https://github.com/torvalds/linux},
}

@misc{FFmpeg,
  author = {FFmpeg},
  year = {2025},
  howpublished  ={[Online]. Available: https://github.com/FFmpeg/FFmpeg},
}

@misc{Git,
  author = {Git},
  year = {2025},
  howpublished  ={[Online]. Available: https://github.com/git/git},
}

@misc{Php,
  author = {Php},
  year = {2025},
  howpublished  ={[Online]. Available: https://github.com/php/php-src},
}

@misc{Libav,
  author = {Libav},
  year = {2025},
  howpublished  ={[Online]. Available: https://github.com/libav/libav},
}

@misc{ImageMagick,
  author = {ImageMagick},
  year = {2025},
  howpublished  ={[Online]. Available: https://github.com/ImageMagick/ImageMagick},
}

@misc{TcpDump,
  author = {TcpDump},
  year = {2025},
  howpublished  ={[Online]. Available: https://github.com/the-tcpdump-group/tcpdump},
}

@misc{Qemu,
  author = {Qemu},
  year = {2025},
  howpublished  ={[Online]. Available: https://github.com/qemu/qemu},
}

@misc{Radare2,
  author = {Radare2},
  year = {2025},
  howpublished  ={[Online]. Available: https://github.com/radareorg/radare2},
}

@misc{Slurm,
  author = {Slurm},
  year = {2025},
  howpublished  ={[Online]. Available: https://github.com/SchedMD/slurm},
}

@misc{cwe,
  author = {Common Weakness Enumerations},
  year = {2025},
  howpublished  ={[Online]. Available: https://cwe.mitre.org},
}

@misc{gpt5,
  author = {GPT-5},
  year = {2025},
  howpublished  ={[Online]. Available: https://openai.com/index/introducing-gpt-5},
}

@misc{idapro,
  author = {IDA Pro},
  year = {2025},
  howpublished  ={[Online]. Available: https://hex-rays.com/ida-pro},
}

@misc{ghidra,
  author = {Ghidra},
  year = {2025},
  howpublished  ={[Online]. Available: https://github.com/NationalSecurityAgency/ghidra},
}

@misc{meta,
  author = {Meta},
  year = {2025},
  howpublished  ={[Online]. Available: https://www.meta.com},
}

@misc{cve201816643,
  author = {CVE-2018-16643},
  year = {2025},
  howpublished  ={[Online]. Available: https://www.cve.org/CVERecord?id=CVE-2018-16643},
}

@misc{cve201713725,
  author = {CVE-2017-13725},
  year = {2025},
  howpublished  ={[Online]. Available: https://www.cve.org/CVERecord?id=CVE-2017-13725},
}

@misc{cve20231801,
  author = {CVE-2023-1801},
  year = {2025},
  howpublished  ={[Online]. Available: https://www.cve.org/CVERecord?id=CVE-2023-1801},
}

@misc{cve201712897,
  author = {CVE-2017-12897},
  year = {2025},
  howpublished  ={[Online]. Available: https://www.cve.org/CVERecord?id=CVE-2017-12897},
}

@misc{cve201713051,
  author = {CVE-2017-13051},
  year = {2025},
  howpublished  ={[Online]. Available: https://www.cve.org/CVERecord?id=CVE-2017-13051},
}

\end{document}